\documentclass[12pt]{iopart}

\usepackage{iopams}
\usepackage{graphicx}
\usepackage{mathrsfs}
\usepackage{color}
\usepackage[table]{xcolor}
\usepackage{hyperref}
\hypersetup{colorlinks, citecolor=black, filecolor=black, linkcolor=black, urlcolor=black}

\expandafter\let\csname equation*\endcsname\relax
\expandafter\let\csname endequation*\endcsname\relax
\usepackage{amsmath}

\definecolor{grayish}{RGB}{230,230,230}

\newcommand{\refEq}[1] {(\ref{#1})}

\newcommand{\Sin}[1]{\ensuremath{\sin \left( #1 \right)}}
\newcommand{\Cos}[1]{\ensuremath{\cos \left( #1 \right)}}

\newcommand{\Exp}[1]{\ensuremath{\exp \left( #1 \right)}}
\newcommand{\tensor}[1]{\ensuremath{\overset{\text{\tiny$\leftrightarrow$}}{#1}}}

\begin{document}

\title[Poloidal tilting symmetry of high order flux surface shaping]{Poloidal tilting symmetry of high order tokamak flux surface shaping in gyrokinetics}

\author{Justin Ball, Felix I. Parra, and Michael Barnes}

\address{Rudolf Peierls Centre for Theoretical Physics, University of Oxford, Oxford OX1 3NP, United Kingdom}
\address{Culham Centre for Fusion Energy, Culham Science Centre, Abingdon OX14 3DB, United Kingdom}
\ead{Justin.Ball@physics.ox.ac.uk}

\begin{abstract}

A poloidal tilting symmetry of the local nonlinear $\delta f$ gyrokinetic model is demonstrated analytically and verified numerically. This symmetry shows that poloidally rotating all the flux surface shaping effects with large poloidal mode number by a single tilt angle has an exponentially small effect on the transport properties of a tokamak. This is shown using a generalization of the Miller local equilibrium model to specify an arbitrary flux surface geometry. With this geometry specification we find that, when performing an expansion in large flux surface shaping mode number, the governing equations of gyrokinetics are symmetric in the poloidal tilt of the high order shaping effects. This allows us to take the fluxes from a single configuration and calculate the fluxes in any configuration that can be produced by tilting the large mode number shaping effects. This creates a distinction between tokamaks with mirror symmetric flux surfaces and tokamaks without mirror symmetry, which is expected to have important consequences for generating toroidal rotation using up-down asymmetry.

\end{abstract}

%\maketitle

%\tableofcontents

%===================================================%
%===================================================%
\section{Introduction}
\label{sec:introduction}
%===================================================%
%===================================================%

Turbulence has been experimentally shown to dominate transport in tokamaks \cite{WoottonTurbReview1990}. In the last 30 years, the fusion community has made remarkable progress in understanding turbulence using gyrokinetics \cite{CattoLinearizedGyrokinetics1978, FriemanNonlinearGyrokinetics1982}. Nonetheless, analytic solutions to the gyrokinetic model are very difficult to find and necessitate many simplifications \cite{RomanelliIonTempCalc1989, RosenbluthPoloidalFlows1998, HintonPoloidalFlows1999, XiaoZFlevel2006}. Typically large, expensive computer simulations are used to find solutions for realistic configurations \cite{DorlandETGturb2000, JenkoGENE2000, CandyGYRO2003, PeetersGKW2009}. However, it is possible to use analytic techniques to establish properties of the gyrokinetic model and constrain possible solutions \cite{PeetersMomTransSym2005, ParraUpDownSym2011, SugamaUpDownSym2011}.

\begin{figure}
 \hspace{0.04\textwidth} (a) \hspace{0.4\textwidth} (b) \hspace{0.25\textwidth}
 \begin{center}
  \includegraphics[width=0.45\textwidth]{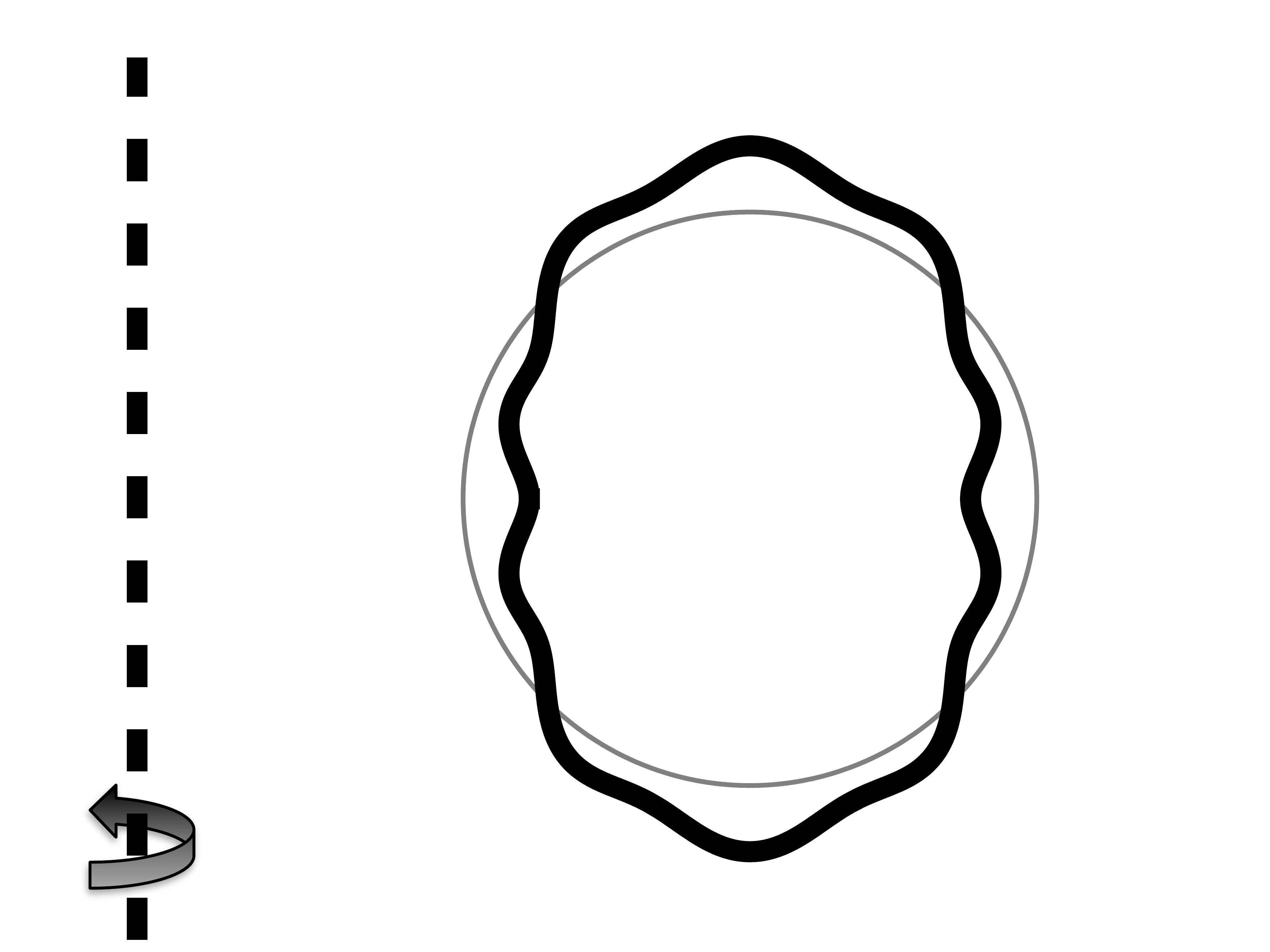}
  \includegraphics[width=0.45\textwidth]{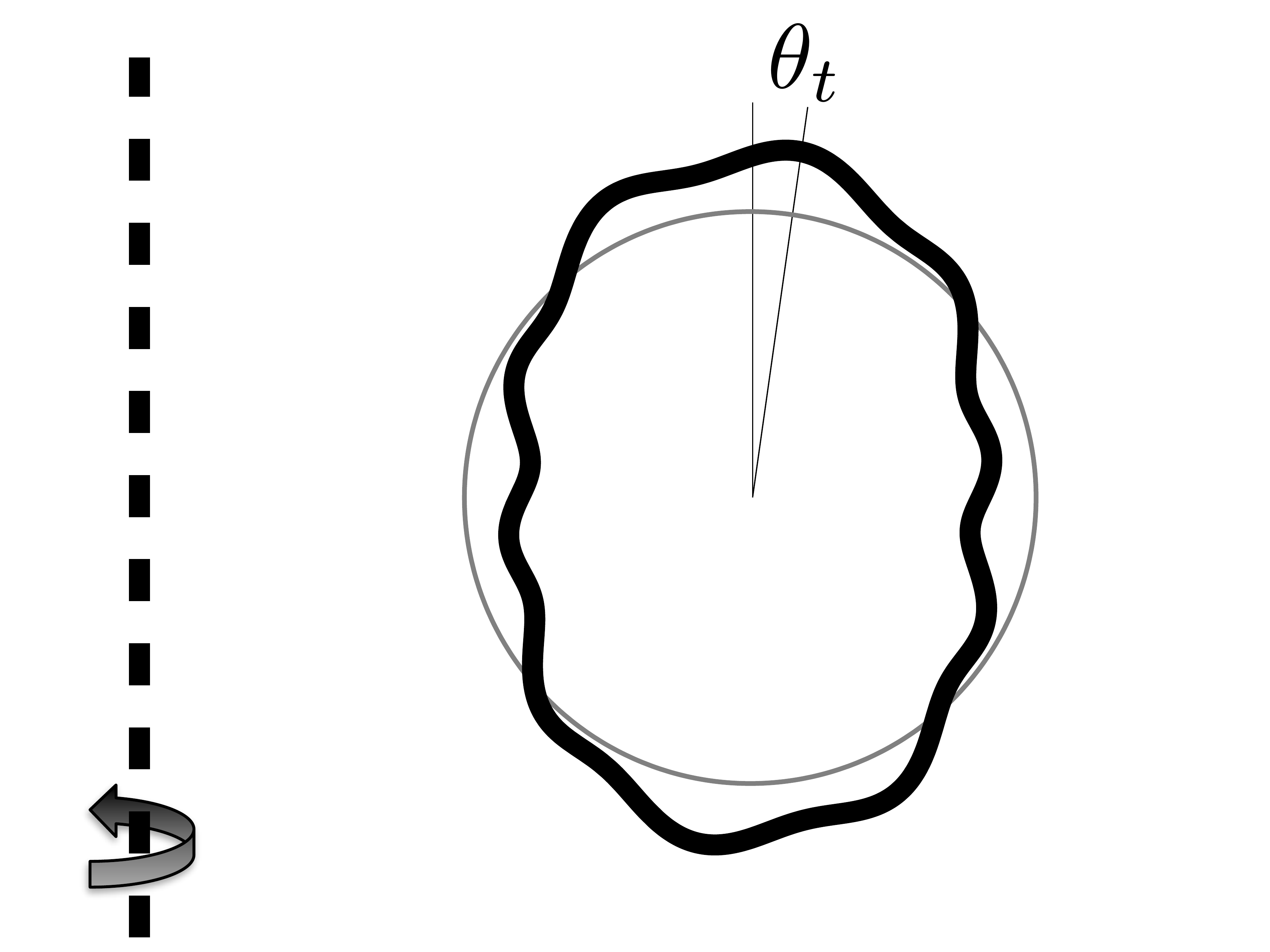}
 \end{center}
 \caption{Example poloidal cross-section of a tokamak (a) without any tilt and (b) with the high order flux surface shaping effects tilted by an angle $\theta_{t}$. Circular flux surfaces are shown in gray for comparison. The axis of axisymmetry is indicated by a dashed line.}
 \label{fig:tiltedGeometry}
\end{figure}

In this work, we show a new symmetry of the gyrokinetic equations. This symmetry means that poloidally rotating all the high order flux surface shaping effects (i.e. shaping effects with a large poloidal mode number) by a single tilt angle does not affect the transport properties of a tokamak (see figure \ref{fig:tiltedGeometry}). To establish this symmetry we expand the high-flow gyrokinetic equations in the limit of high order flux surface shaping mode number. We expect turbulent eddies to extend along the field line and average over the very rapid variation created by high mode number shaping. Therefore, it is intuitive that the effect of tilting flux surface shaping should diminish in the limit of high order shaping. However, what is surprising is that this symmetry proves that the effect diminishes exponentially with mode number, rather than polynomially. Hence we find that tilting high order flux surface shaping has an exponentially small effect on the turbulent fluxes. We will see that this is particularly relevant because of the recent interest in up-down asymmetric tokamak geometries (i.e. tokamaks that are not mirror symmetric about the midplane) \cite{CamenenPRLExp2010, BallMomUpDownAsym2014}.

Breaking the up-down symmetry of the flux surfaces is potentially beneficial \cite{CamenenPRLExp2010, BallMomUpDownAsym2014} because it removes a constraint that limits turbulent momentum transport to be small in $\rho_{\ast} \equiv \rho_{i} / a \ll 1$, where $\rho_{i}$ is the ion gyroradius and $a$ is the tokamak minor radius \cite{ParraIntrinsicRotReview2012}.  In fact, reference \cite{CamenenPRLExp2010} presents results from the TCV tokamak that have provided the first experimental evidence of intrinsic rotation generated by up-down asymmetry. These experiments used up-down asymmetric flux surfaces to change the rotation profile by between 50\% and 100\%. This motivates further investigation of up-down asymmetric devices because significant momentum transport has been shown to improve tokamak performance. Experiments on DIII-D \cite{StraitExpRWMstabilizationD3D1995, GarofaloExpRWMstabilizationD3D2002} and TEXTOR \cite{deVriesRotMHDStabilization1996} (among others) used plasma rotation to stabilize resistive wall modes, a dangerous MHD instability, and enable discharges with plasma beta that exceeded the Troyon limit \cite{TroyonMHDLimit1984}. Additionally, a gradient in rotation is theoretically expected to decorrelate turbulence and directly reduce energy transport \cite{BarnesFlowShear2011}. Reference \cite{BurrellShearTurbStabilization1997} observes that the best fusion performance in the DIII-D and JT-60U tokamaks has been obtained under conditions where turbulence reduction by velocity shear is almost certainly taking place.

However, the space of up-down asymmetric flux surface shapes is enormous. One class of shapes is those that can be produced by tilting an up-down symmetric configuration poloidally by a single tilt angle. This makes flux surfaces that still have mirror symmetry, but no longer have up-down symmetry. The symmetry presented in this work shows that, because of the constraint on momentum transport in up-down symmetric devices \cite{PeetersMomTransSym2005, ParraUpDownSym2011, BallMomUpDownAsym2014, CamenenPRLExp2010}, we would expect these mirror symmetric configurations to generate little momentum flux, in the limit of high order shaping effects. Consequently, this establishes a distinction between devices with mirror symmetric flux surfaces and devices without mirror symmetry, which may have important consequences for flux surface shaping of any mode number. Additionally, the exponential scaling suggests that generating rotation using up-down asymmetric triangularity or squareness will be less effective than up-down asymmetric elongation. The approximate symmetry presented in this paper indicates that the geometry used in the TCV up-down asymmetry experiments \cite{CamenenPRLExp2010} is close to the optimal mirror symmetric shape for generating large rotation, but this has not yet been tested experimentally. Regardless, a significant enhancement over the TCV results may still be found in the space of non-mirror symmetric shapes.

While the symmetry presented in this work has the biggest practical implication for momentum transport, it also applies to energy and particle transport. For example, references \cite{WeisenShapeOnConfinement1997, CamenenTriConfinement2007} look at the effect of elongation and triangularity on the energy confinement time in TCV. From the scaling in our work, we would expect that tilting triangularity or higher order shaping would have a smaller effect on energy confinement compared with tilting elongation. This can have significance for purely up-down symmetric configurations. For example, horizontal elongation can be thought of as vertical elongation with a $90^{\circ}$ tilt just as changing the sign of triangularity is equivalent to tilting the triangularity by $180^{\circ}$. Therefore, we would expect switching from vertical to horizontal elongation would have a larger effect on the energy confinement time than changing the sign of the triangularity.

Section \ref{sec:analyticArg} of this paper contains the analytic analysis, which includes introducing gyrokinetics, detailing a generalized version of the Miller local equilibrium specification, and demonstrating the poloidal tilting symmetry of high order flux surface shaping. Section \ref{sec:numResults} presents the results of nonlinear local gyrokinetic simulations. These simulations are aimed at providing numerical verification of the analytic work. Finally, section \ref{sec:conclusions} offers a summary and some concluding remarks.

%===================================================%
%===================================================%
\section{Poloidal tilting symmetry of high order flux surface shaping}
\label{sec:analyticArg}
%===================================================%
%===================================================%

In this section we will show the tilting symmetry of high order flux surface shaping in the gyrokinetic model. First we will give the governing equations of the complete nonlinear local $\delta f$ gyrokinetic model, including electromagnetic effects and rotation. In these equations we will find several geometric coefficients that must be calculated from the tokamak equilibrium. To do so, we will generalize the traditional Miller local equilibrium \cite{MillerGeometry1998} to specify arbitrarily shaped flux surfaces using Fourier analysis. This reveals how the effect of high order shaping enters into the geometric coefficients and hence the gyrokinetic model. Finally we will expand the gyrokinetic equations in the limit of high order flux surface shaping and show that tilting high order shaping does not affect particle, momentum, or energy transport.

%===================================================%
\subsection{Gyrokinetics}
\label{subsec:gyrokinetics}
%===================================================%

Gyrokinetics has many variations \cite{LeeGenFreqGyro1983, LeeParticleSimGyro1983, DubinHamiltonianGyro1983, HahmGyrokinetics1988, SugamaGyroTransport1996, SugamaHighFlowGyro1998, BrizardGyroFoundations2007, ParraGyrokineticLimitations2008, ParraLagrangianGyro2011, AbelGyrokineticsDeriv2012}, but is based on the expansion of the Fokker-Plank and Maxwell's equations in $\rho_{\ast} \equiv \rho_{i} / a \ll 1$. This model investigates plasma behavior with timescales much slower than the ion gyrofrequency $\Omega_{i}$ and the electron gyrofrequency $\Omega_{e}$ (i.e. $\omega \ll \Omega_{i} \ll \Omega_{e}$), but retains the finite size of the gyroradius by assuming that the perpendicular wavenumber of the turbulence is comparable to the ion gyroradius (i.e. $k_{\perp} \rho_{i} \sim 1$ where $k_{\perp}$ is the characteristic wavenumber of the turbulence perpendicular to the magnetic field). In this limit, the six dimensions of velocity space reduce to five because the particle gyrophase can be ignored. As such, gyrokinetics evolves rings of charge as they generate and respond to electric and magnetic fields. In this paper we will use $\delta f$ gyrokinetics, which assumes that the turbulence arises from perturbations to the distribution function that are small compared to the background (i.e. $f_{s 1} \ll f_{s 0}$, where $f_{s 0}$ is the background distribution function for species $s$ and $f_{s 1}$ is the lowest order perturbation). These particular choices have been shown experimentally to be appropriate for modeling core turbulence \cite{McKeeTurbulenceScale2001}. Furthermore, we will assume the plasma is sufficiently collisional so that the background distribution function is Maxwellian, 
\begin{align}
   f_{s 0} = F_{M s} \equiv n_{s} \left( \frac{m_{s}}{2 \pi T_{s}} \right)^{3/2} \text{exp} \left( - \frac{m_{s} w^{2}}{2 T_{s}} \right) . \label{eq:maxwellianDef}
\end{align}
Here $n_{s}$ is the density of species $s$, $m_{s}$ is the particle mass, $T_{s}$ is the temperature, $\vec{w} \equiv \vec{v} - R \Omega_{\zeta} \hat{e}_{\zeta}$ is the velocity shifted into the rotating frame, $R$ is the major radial coordinate, $\Omega_{\zeta}$ is the toroidal rotation frequency, $\zeta$ is the toroidal angle, and $\hat{e}_{\zeta}$ is the unit vector in the toroidal direction. To lowest order in $\rho_{i} / a \ll 1$, it can be shown that all species rotate at $\Omega_{\zeta} = - d \Phi_{-1} / d \psi$, where $\Phi_{-1} \sim \rho_{\ast}^{-1} T_{e} / e$ is the lowest order electrostatic potential and a flux function \cite{HintonNeoclassicalRotation1985, ConnorMomTransport1987, CattoRotation1987}. Here $\psi$ is the poloidal magnetic flux and $e$ is the proton electric charge. While $T_{s}$ and $\Omega_{\zeta}$ are flux functions, we note that (due to the centrifugal force) density is not a flux function, but is instead given by \cite{MaschkeEquilWithRotation1980}
\begin{align}
   n_{s} \left( \psi, \theta \right) = \eta_{s} \left( \psi \right) \Exp{ \frac{m_{s} R^{2} \Omega_{\zeta}^{2}}{2 T_{s}} - \frac{Z_{s} e \Phi_{0}}{T_{s}}} ,
\end{align}
where $\eta_{s} \left( \psi \right)$ is the pseudo-density flux function, $\theta$ is a poloidal angle, $Z_{s}$ is the electric charge number, and $\Phi_{0}$ is the next order electrostatic potential. We can find $\Phi_{0}$ by imposing quasineutrality,
\begin{align}
   \sum_{s} Z_{s} e n_{s} = \sum_{s} Z_{s} e \eta_{s} \left( \psi \right) \Exp{ \frac{m_{s} R^{2} \Omega_{\zeta}^{2}}{2 T_{s}} - \frac{Z_{s} e \Phi_{0}}{T_{s}}} = 0 . \label{eq:backgroundQuasi}
\end{align}

From the assumption that $k_{\perp} \rho_{i} \sim 1$ (remembering our expansion in $\rho_{i} / a \ll 1$), we know that the background plasma quantities vary little on the scale of the turbulence in the directions perpendicular to the background magnetic field. This is called the local approximation and it motivates using periodic boundary conditions in the perpendicular directions. Ballooning coordinates \cite{BeerBallooingCoordinates1995} are generally used in local gyrokinetics to model turbulence in a flux tube, a long narrow domain that follows a single field line. These boundary conditions allow us to Fourier analyze in the poloidal flux $\psi$ (which parameterizes the radial direction) and in
\begin{align}
   \alpha \equiv \zeta - I \left( \psi \right) \left. \int_{\theta_{\alpha} \left( \psi \right)}^{\theta} \right|_{\psi} d \theta' \left( R^{2} \vec{B} \cdot \vec{\nabla} \theta' \right)^{-1} - \Omega_{\zeta} t \label{eq:alphaDef}
\end{align}
(which parameterizes the direction perpendicular to the field lines, but within the flux surface). Here $I \left( \psi \right) \equiv R B_{\zeta}$ is the toroidal field flux function, $\vec{B}$ is the background magnetic field (which we require to be axisymmetric), and $t$ is the time. Note the free parameter $\theta_{\alpha} \left( \psi \right)$, which determines the field line selected by $\alpha = 0$ on each flux surface and will be important later in this work.

The high-flow, Fourier analyzed gyrokinetic equation can be written as \cite{ParraUpDownSym2011}
\begin{align}
   \frac{\partial h_{s}}{\partial t} &+ w_{||} \hat{b} \cdot \vec{\nabla} \theta \left. \frac{\partial h_{s}}{\partial \theta} \right|_{w_{||}, \mu} + i \left( k_{\psi} v_{d s \psi} + k_{\alpha} v_{d s \alpha} \right) h_{s} + a_{s ||} \left. \frac{\partial h_{s}}{\partial w_{||}} \right|_{\theta, \mu} - \sum_{s'} \langle C_{ss'}^{\left( l \right)} \rangle_{\varphi} \nonumber \\
&+ \left\{ \langle \chi \rangle_{\varphi}, h_{s} \right\} = \frac{Z_{s} e F_{M s}}{T_{s}} \frac{\partial \langle \chi \rangle_{\varphi}}{\partial t} - v_{\chi s \psi} F_{M s} \left[ \frac{1}{n_{s}} \left. \frac{\partial n_{s}}{\partial \psi} \right|_{\theta} \right. \label{eq:gyrokineticEq} \\
 &+ \left. \frac{m_{s} I w_{||}}{B T_{s}} \frac{d \Omega_{\zeta}}{d \psi} + \frac{Z_{s} e}{T_{s}} \left. \frac{\partial \Phi_{0}}{\partial \psi} \right|_{\theta} - \frac{m_{s} R \Omega_{\zeta}^{2}}{T_{s}} \left. \frac{\partial R}{\partial \psi} \right|_{\theta} + \left( \frac{m_{s} w^{2}}{2 T_{s}} - \frac{3}{2} \right) \frac{1}{T_{s}} \frac{d T_{s}}{d \psi} \right] , \nonumber
\end{align}
where the coordinates are $t$ (the time), $\theta$ (a poloidal angle), $k_{\psi}$ (the radial wavenumber), $k_{\alpha}$ (the poloidal wavenumber), $w_{||}$ (the parallel velocity in the rotating frame), $\mu \equiv m_{s} w_{\perp}^{2} / 2 B$ (the magnetic moment), and we have already eliminated $\varphi$ (the gyrophase) by gyroaveraging. The unknowns are
\begin{align}
   h_{s} \equiv \left\langle \left\langle \left( f_{s 1} + \frac{Z_{s} e \phi}{T_{s}} F_{M s} \right) \Exp{-i k_{\psi} \psi - i k_{\alpha} \alpha} \right\rangle_{\Delta \psi} \right\rangle_{\Delta \alpha} \label{eq:distFnFourierAnalysis}
\end{align}
(the Fourier analyzed nonadiabatic portion of the distribution function) and the fields contained in
\begin{align}
   \left\langle \chi \right\rangle_{\varphi} \equiv J_{0} \left( k_{\perp} \rho_{s} \right) \left( \phi - w_{||} A_{||} \right) + \frac{1}{\Omega_{s}} \frac{\mu B}{m_{s}} \frac{2 J_{1} \left( k_{\perp} \rho_{s} \right)}{k_{\perp} \rho_{s}} B_{||}
\end{align}
(the Fourier analyzed gyroaveraged generalized potential). Here $\left\langle \ldots \right\rangle_{\Delta \psi} \equiv \Delta \psi^{-1} \int_{\Delta \psi} d \psi \left( \ldots \right)$ is a coarse-grain average over the radial distance $\Delta \psi$ (which is larger then the scale of the turbulence, but smaller than the scale of the device), $\left\langle \ldots \right\rangle_{\Delta \alpha} \equiv \Delta \alpha^{-1} \int_{\Delta \alpha} d \alpha \left( \ldots \right)$ is a coarse-grain average over the poloidal distance $\Delta \alpha$ (which is larger then the scale of the turbulence, but smaller than the scale of the device), $\left\langle \ldots \right\rangle_{\varphi}$ is the gyroaverage at fixed guiding center, $J_{n} \left( \ldots \right)$ is the $n$th order Bessel function of the first kind, $\phi$ is the Fourier analyzed perturbed electrostatic potential, $A_{||}$ is the Fourier analyzed perturbed magnetic vector potential, $B_{||}$ is the component of the Fourier analyzed perturbed magnetic field parallel to the background magnetic field,
\begin{align}
   k_{\perp} = \sqrt{k_{\psi}^{2} \left| \vec{\nabla} \psi \right|^{2} + 2 k_{\psi} k_{\alpha} \vec{\nabla} \psi \cdot \vec{\nabla} \alpha + k_{\alpha}^{2} \left| \vec{\nabla} \alpha \right|^{2}} \label{eq:kperpDef}
\end{align}
is the perpendicular wavevector, $\rho_{s} \equiv \sqrt{2 \mu B / m_{s}} / \Omega_{s}$ is the gyroradius, $\Omega_{s} \equiv Z_{s} e B / m_{s}$ is the gyrofrequency, and $Z_{s}$ is the species charge number.

The drift coefficients are given by
\begin{align}
   v_{d s \psi} &\equiv \vec{v}_{d s} \cdot \vec{\nabla} \psi \label{eq:radialGCdriftvelocity} \\
   &= \left( - \frac{I}{B} \frac{\partial \Phi_{0}}{\partial \theta} - \frac{I \left( m_{s} w_{||}^{2} + \mu B \right)}{m_{s} \Omega_{s} B} \frac{\partial B}{\partial \theta} + \frac{2 B R \Omega_{\zeta} w_{||}}{\Omega_{s}} \frac{\partial R}{\partial \theta} + \frac{I R \Omega_{\zeta}^{2}}{\Omega_{s}} \frac{\partial R}{\partial \theta} \right) \hat{b} \cdot \vec{\nabla} \theta \nonumber
\end{align}
and
\begin{align}
   v_{d s \alpha} &\equiv \vec{v}_{d s} \cdot \vec{\nabla} \alpha = - \frac{\partial \Phi_{0}}{\partial \psi} + \frac{\partial \Phi_{0}}{\partial \theta} \frac{\hat{b} \cdot \left( \vec{\nabla} \theta \times \vec{\nabla} \alpha \right)}{B} \nonumber \\
   &- \frac{m_{s} w_{||}^{2} + \mu B}{m_{s} \Omega_{s}} \left( \frac{\partial B}{\partial \psi} - \frac{\partial B}{\partial \theta} \frac{\hat{b} \cdot \left( \vec{\nabla} \theta \times \vec{\nabla} \alpha \right)}{B} \right) - \frac{\mu_{0} w_{||}^{2}}{B \Omega_{s}} \left. \frac{\partial p}{\partial \psi} \right|_{R} \label{eq:alphaGCdriftvelocity} \\
   &+ \frac{2 \Omega_{\zeta} w_{||}}{\Omega_{s}} \hat{e}_{\zeta} \cdot \left( \vec{\nabla} \alpha \times \vec{\nabla} R \right) + \frac{m_{s} R \Omega_{\zeta}^{2}}{Z_{s} e} \left( \frac{\partial R}{\partial \psi} - \frac{\partial R}{\partial \theta} \frac{\hat{b} \cdot \left( \vec{\nabla} \theta \times \vec{\nabla} \alpha \right)}{B} \right) , \nonumber
\end{align}
where $\hat{b} \equiv \vec{B} / B$ is the magnetic field unit vector, $\mu_{0}$ is the permeability of free space, $p \equiv \sum_{s} n_{s} T_{s}$ is the plasma pressure, and
\begin{align}
   \left. \frac{\partial p}{\partial \psi} \right|_{R} = \left. \frac{\partial p}{\partial \psi} \right|_{\theta} - \sum_{s} n_{s} m_{s} R \Omega_{\zeta}^{2} \left. \frac{\partial R}{\partial \psi} \right|_{\theta} .
\end{align}
The parallel acceleration is given by
\begin{align}
   a_{s ||} = \left( - \frac{\mu}{m_{s}} \frac{\partial B}{\partial \theta} - \frac{Z_{s} e}{m_{s}} \frac{\partial \Phi_{0}}{\partial \theta} + R \Omega_{\zeta}^{2} \frac{\partial R}{\partial \theta} \right) \hat{b} \cdot \vec{\nabla} \theta , \label{eq:parallelAcceleration}
\end{align}
$C_{s s'}^{\left( l \right)}$ is the linearized collision operator, the nonlinear term is
\begin{align}
  \left\{ \langle \chi \rangle_{\varphi}, h_{s} \right\} = \sum_{k_{\psi}', k_{\alpha}'} \left( k_{\psi}' k_{\alpha} - k_{\psi} k_{\alpha}' \right) \left\langle \chi \right\rangle_{\varphi} \left( k_{\psi}', k_{\alpha}' \right) h_{s} \left( k_{\psi} - k_{\psi}', k_{\alpha} - k_{\alpha}' \right) , \label{eq:nonlinearTerm}
\end{align}
and
\begin{align}
   v_{\chi s \psi} \equiv i k_{\alpha} \left\langle \chi \right\rangle_{\varphi} .
\end{align}

In order to solve for $\phi$, $A_{||}$, and $B_{||}$ we also need the Fourier analyzed quasineutrality equation \cite{ParraUpDownSym2011}
\begin{align}
      \phi = 2 \pi \left( \sum_{s} \frac{Z_{s}^{2} e^{2} n_{s}}{T_{s}} \right)^{-1} \sum_{s} \frac{Z_{s} e B}{m_{s}} \int dw_{||} \int d \mu J_{0} \left( k_{\perp} \rho_{s} \right) h_{s} , \label{eq:quasineut}
\end{align}
parallel current equation \cite{ParraUpDownSym2011}
\begin{align}
      A_{||} = \frac{2 \pi \mu_{0}}{k_{\perp}^{2}} \sum_{s} \frac{Z_{s} e B}{m_{s}} \int dw_{||}  \int d \mu J_{0} \left( k_{\perp} \rho_{s} \right) w_{||} h_{s} , \label{eq:parCur}
\end{align}
and perpendicular current equation \cite{ParraUpDownSym2011}
\begin{align}
      B_{||} = -2 \pi \mu_{0} \sum_{s} \frac{B}{m_{s}} \int dw_{||} \int d \mu \frac{2 J_{1} \left( k_{\perp} \rho_{s} \right)}{k_{\perp} \rho_{s}} \mu h_{s} . \label{eq:perpCur}
\end{align}
Equations \refEq{eq:gyrokineticEq}, \refEq{eq:quasineut}, \refEq{eq:parCur}, and \refEq{eq:perpCur} comprise the nonlinear electromagnetic gyrokinetic model, in the presence of rotation.

Solving the gyrokinetic model for $h_{s}$, $\phi$, $A_{||}$, and $B_{||}$ allows us to calculate the turbulent fluxes of particles, momentum, and energy as well as the turbulent energy exchange between species. The full expressions are written in \ref{app:fluxes}. Here we give only the electrostatic contribution to the particle flux
\begin{align}
   \Gamma_{s}^{\phi} &\equiv - \left\langle R \left\langle \left\langle \int d^{3} w \overline{h}_{s} \hat{e}_{\zeta} \cdot \delta \vec{E} \right\rangle_{\Delta \psi} \right\rangle_{\Delta t} \right\rangle_{\psi} \label{eq:partFluxDef} \\
  &= \frac{4 \pi^{2} i}{m_{s} V'} \left\langle \sum_{k_{\psi}, k_{\alpha}} k_{\alpha} \oint d \theta J B \phi \left( k_{\psi}, k_{\alpha} \right) \int dw_{||} d \mu ~ h_{s} \left( - k_{\psi}, - k_{\alpha} \right) J_{0} \left( k_{\perp} \rho_{s} \right) \right\rangle_{\Delta t} , \label{eq:partFlux}
\end{align}
the momentum flux
\begin{align}
   \Pi_{s}^{\phi} &\equiv - \left\langle R \left\langle \left\langle \int d^{3} w \overline{h}_{s} m_{s} R \left( \vec{w} \cdot \hat{e}_{\zeta} + R \Omega_{\zeta} \right) \hat{e}_{\zeta} \cdot \delta \vec{E} \right\rangle_{\Delta \psi} \right\rangle_{\Delta t} \right\rangle_{\psi} \label{eq:momFluxDef} \\
  &= \frac{4 \pi^{2} i}{V'} \left\langle \sum_{k_{\psi}, k_{\alpha}} k_{\alpha} \oint d \theta J B \phi \left( k_{\psi}, k_{\alpha} \right) \int dw_{||} d \mu ~ h_{s} \left( - k_{\psi}, - k_{\alpha} \right) \right. \label{eq:momFlux} \\
   &\times \left. \left[ \left( \frac{I}{B} w_{||} + R^{2} \Omega_{\zeta} \right) J_{0} \left( k_{\perp} \rho_{s} \right) + \frac{i}{\Omega_{s}} \frac{k^{\psi}}{B} \frac{\mu B}{m_{s}} \frac{2 J_{1} \left( k_{\perp} \rho_{s} \right)}{k_{\perp} \rho_{s}} \right] \right\rangle_{\Delta t} , \nonumber
\end{align}
the energy flux
\begin{align}
   Q_{s}^{\phi} &\equiv - \left\langle R \left\langle \left\langle \int d^{3} w \overline{h}_{s} \left( \frac{m_{s}}{2} w^{2} + Z_{s} e \Phi_{0} - \frac{m_{s}}{2} R^{2} \Omega_{\zeta}^{2} \right) \hat{e}_{\zeta} \cdot \delta \vec{E} \right\rangle_{\Delta \psi} \right\rangle_{\Delta t} \right\rangle_{\psi} \label{eq:heatFluxDef} \\
  &= \frac{4 \pi^{2} i}{V'} \left\langle \sum_{k_{\psi}, k_{\alpha}} k_{\alpha} \oint d \theta J B \phi \left( k_{\psi}, k_{\alpha} \right) \int dw_{||} d \mu ~ h_{s} \left( - k_{\psi}, - k_{\alpha} \right) \right. \label{eq:heatFlux} \\
   &\times \left. \left( \frac{w^{2}}{2} + \frac{Z_{s} e \Phi_{0}}{m_{s}} - \frac{1}{2} R^{2} \Omega_{\zeta}^{2} \right) J_{0} \left( k_{\perp} \rho_{s} \right) \right\rangle_{\Delta t} , \nonumber
\end{align}
and the turbulent energy exchange between species
\begin{align}
   P_{Q s}^{\phi} &\equiv \left\langle \left\langle \left\langle \int d^{3} w Z_{s} e \overline{h}_{s} \frac{\partial \overline{\phi}}{\partial t} \right\rangle_{\Delta \psi} \right\rangle_{\Delta t} \right\rangle_{\psi} \label{eq:heatingDef} \\
   &= \frac{4 \pi^{2}}{V'} \left\langle \sum_{k_{\psi}, k_{\alpha}} \oint d \theta J \Omega_{s} \frac{\partial}{\partial t} \left( \phi \left( k_{\psi}, k_{\alpha} \right) \right) \int dw_{||} d \mu ~ h_{s} \left( - k_{\psi}, - k_{\alpha} \right) J_{0} \left( k_{\perp} \rho_{s} \right) \right\rangle_{\Delta t} . \label{eq:heating}
\end{align}
Here $\overline{h}_{s} \equiv f_{s 1} + Z_{s} e \overline{\phi} F_{M s} / T_{s}$ is the nonadiabatic portion of the distribution function, $\overline{\left( \ldots \right)}$ indicates the quantity has not been Fourier analyzed, $\delta \vec{E} = - \vec{\nabla}_{\perp} \overline{\phi}$ is the turbulent electric field, $\left\langle \ldots \right\rangle_{\psi} \equiv \left( 2 \pi / V' \right) \oint_{0}^{2 \pi} d \theta J \left( \ldots \right)$ is the flux surface average, $\left\langle \ldots \right\rangle_{\Delta t} \equiv \Delta t^{-1} \int_{\Delta t} dt \left( \ldots \right)$ is a coarse-grain average over a time $\Delta t$ (which is longer than the turbulent decorrelation time), $V' \equiv 2 \pi \oint d\theta J$, $J \equiv \left| \vec{B} \cdot \vec{\nabla} \theta \right|^{-1}$ is the Jacobian, and $k^{\psi} \equiv \vec{k}_{\perp} \cdot \vec{\nabla} \psi = k_{\psi} \left| \vec{\nabla} \psi \right|^{2} + k_{\alpha} \vec{\nabla} \psi \cdot \vec{\nabla} \alpha$.

In this work we are concerned with the effect of geometry on the turbulent fluxes. All of the information concerning the tokamak geometry enters the gyrokinetic model via ten geometric coefficients: $B$, $\hat{b} \cdot \vec{\nabla} \theta$, $v_{d s \psi}$, $v_{d s \alpha}$, $a_{s ||}$, $\left| \vec{\nabla} \psi \right|^{2}$, $\vec{\nabla} \psi \cdot \vec{\nabla} \alpha$, $\left| \vec{\nabla} \alpha \right|^{2}$, $R$, and $\left. \partial R / \partial \psi \right|_{\theta}$. In order to calculate these geometric coefficients we will first need to specify the background plasma equilibrium.

%===================================================%
\subsection{Miller local equilibrium specification}
\label{subsec:millerGeo}
%===================================================%

\begin{figure}
 \begin{center}
  \includegraphics[width=0.4\textwidth]{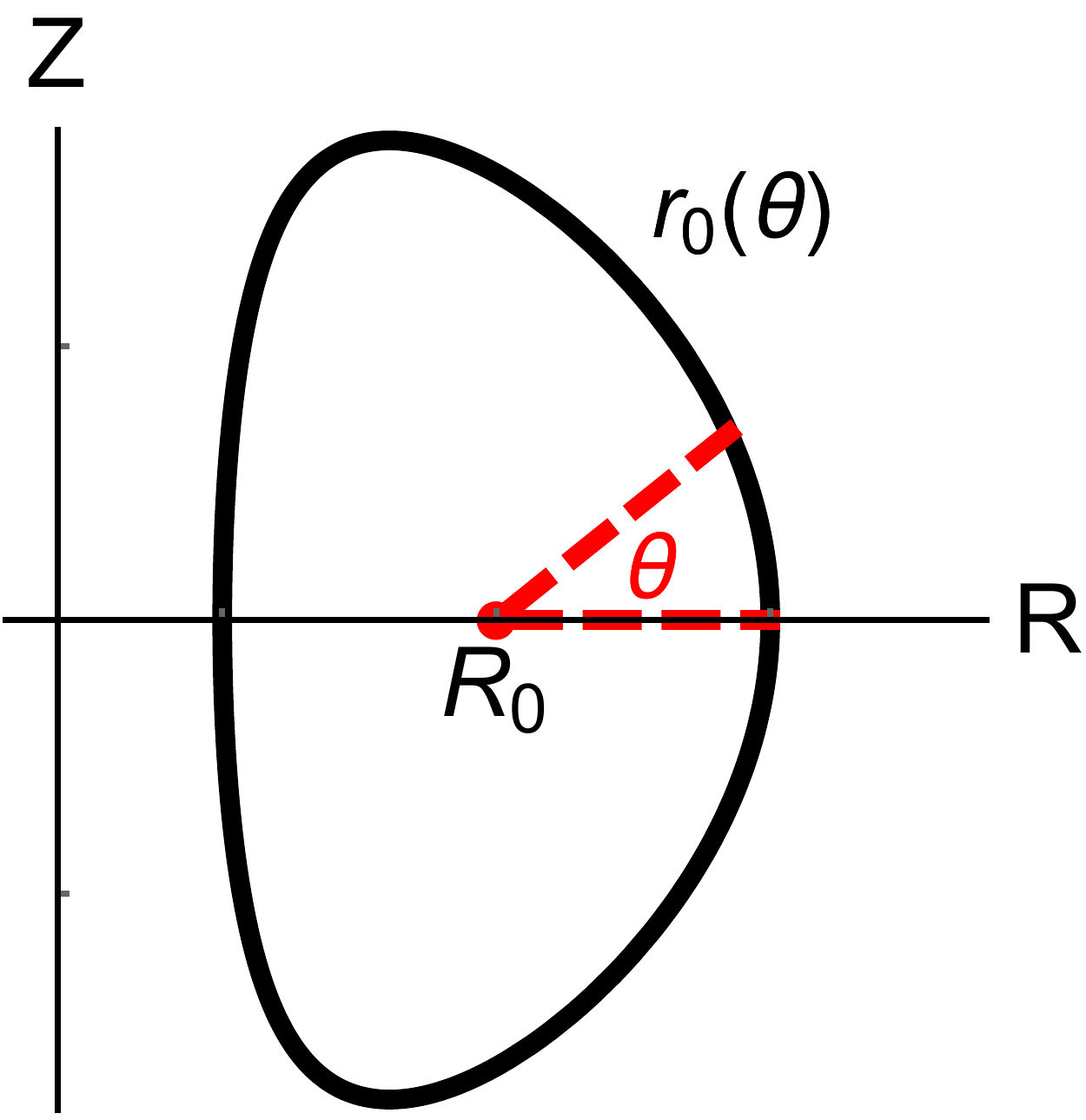}
 \end{center}
 \caption{An example flux surface of interest, $r_{0} \left( \theta \right)$, needed by equation \refEq{eq:gradShafranovLocalEq} for the Miller local equilibrium model.}
 \label{fig:coordinateSystem}
\end{figure}

To specify the background tokamak equilibrium for our local gyrokinetic model we will use a generalization of the Miller local equilibrium model \cite{MillerGeometry1998}. The Miller prescription approximates the equilibrium around a single flux surface of interest when given: $R_{0}$ (the tokamak major radius), $r_{0} \left( \theta \right)$ (the shape of the flux surface of interest), $\left. \partial r / \partial r_{\psi} \right|_{r_{\psi 0}, \theta}$ (how the shape of the flux surface of interest changes with minor radius), and (in the absence of significant rotation) four scalar quantities. This information can be used to construct the nearby flux surfaces according to
\begin{align}
   r \left( r_{\psi}, \theta \right) &= r_{0} \left( \theta \right) + \left. \frac{\partial r}{\partial r_{\psi}} \right|_{r_{\psi 0}, \theta} \left( r_{\psi} - r_{\psi 0} \right) \label{eq:gradShafranovLocalEq} \\
   R \left( r_{\psi}, \theta \right) &= R_{0} + r \left( r_{\psi}, \theta \right) \Cos{\theta} \label{eq:geoMajorRadius} \\
   Z \left( r_{\psi}, \theta \right) &= r \left( r_{\psi}, \theta \right) \Sin{\theta} , \label{eq:geoAxial}
\end{align}
where $r_{\psi}$ is a minor radial coordinate and $r_{\psi 0}$ is the minor radial location of the flux surface of interest (see figure \ref{fig:coordinateSystem}).
The four scalar quantities are commonly taken to be $I$ (the toroidal field flux function), $q$ (the safety factor), $dq/dr_{\psi}$ (the magnetic shear), and $d p / d r_{\psi}$ (the pressure gradient) of the flux surface of interest. However, when the plasma is rotating quickly, $d p / d r_{\psi}$ is replaced by $\left. \partial p / \partial r_{\psi} \right|_{R}$, which requires four additional, species-dependent parameters: $\eta_{s} T_{s}$ (the pseudo-pressure), $d \left( \eta_{s} T_{s} \right) / d r_{\psi}$ (the derivative of the pseudo-pressure), $m_{s} \Omega_{\zeta}^{2} / 2 T_{s}$ (a rotational frequency parameter), and $d \left( m_{s} \Omega_{\zeta}^{2} / 2 T_{s} \right) / d r_{\psi}$ (the derivative of the rotational frequency parameter).

The functions $r_{0} \left( \theta \right)$ and $\left. \partial r / \partial r_{\psi} \right|_{r_{\psi 0}, \theta}$ allow us to calculate poloidal derivatives of any order as well as the first order radial derivatives: $\left. \partial R / \partial r_{\psi} \right|_{\theta}$ and $\left. \partial Z / \partial r_{\psi} \right|_{\theta}$ (but not higher order radial derivatives). This is enough information to calculate the poloidal magnetic field through
\begin{align}
   \vec{B}_{p} = \vec{\nabla} \zeta \times \vec{\nabla} r_{\psi} \frac{d \psi}{d r_{\psi}} ,
\end{align}
where we can use the identity
\begin{align}
   \vec{\nabla} u_{1} &= \frac{\partial \vec{r} / \partial u_{2} \times \partial \vec{r} / \partial u_{3}}{\partial \vec{r} / \partial u_{1} \cdot \left( \partial \vec{r} / \partial u_{2} \times \partial \vec{r} / \partial u_{3} \right)} \label{eq:gradIdentity}
\end{align}
for $\left( u_{1}, u_{2}, u_{3} \right)$, a cyclic permutation of $\left( r_{\psi}, \theta, \zeta \right)$. Note that we can determine $d \psi / d r_{\psi}$ from $q$ using
\begin{align}
   q \equiv \frac{I}{2 \pi} \left. \oint_{0}^{2 \pi} \right|_{\psi} d \theta \left( R^{2} \vec{B}_{p} \cdot \vec{\nabla} \theta \right)^{-1} . \label{eq:safetyFactorDef}
\end{align}
In addition to giving $\vec{B}_{p}$, the derivative $d \psi / d r_{\psi}$ is used to calculate $d / d \psi$ from $d / d r_{\psi}$. 

To calculate some of the geometric coefficients (e.g. those containing $\vec{\nabla} \alpha$), we will need to take the radial derivative of $B_{p} \equiv \left| \vec{B}_{p} \right|$. However, in the local Miller model we cannot calculate this directly from our geometry specification because it contains second order radial derivatives. Instead we calculate it by ensuring that the Grad-Shafranov equation \cite{GradGradShafranovEq1958},
\begin{align}
   R^{2} \vec{\nabla} \cdot \left( \frac{\vec{\nabla} \psi}{R^{2}} \right) = -\mu_{0} R^{2} \left. \frac{\partial p}{\partial \psi} \right|_{R} - I \frac{d I}{d \psi} , \label{eq:gradShafranov}
\end{align}
is satisfied. Using the Grad-Shafranov equation,
\begin{align}
   R^{2} \vec{\nabla} \cdot \left( \frac{\vec{\nabla} \psi}{R^{2}} \right) &= \frac{R^{2}}{J} \left. \frac{\partial}{\partial \psi} \right|_{\theta} \left( \frac{J}{R^{2}} \left| \vec{\nabla} \psi \right|^{2} \right) + \frac{R^{2}}{J} \left. \frac{\partial}{\partial \theta} \right|_{\psi} \left( \frac{J}{R^{2}} \vec{\nabla} \psi \cdot \vec{\nabla} \theta \right) ,
\end{align}
equation \refEq{eq:gradIdentity}, and the Jacobian
\begin{align}
   J &\equiv \left| \vec{\nabla} \psi \cdot \left( \vec{\nabla} \theta \times \vec{\nabla} \zeta \right) \right|^{-1} = \left( \vec{B}_{p} \cdot \vec{\nabla} \theta \right)^{-1} = \frac{1}{B_{p}} \left. \frac{\partial l_{p}}{\partial \theta} \right|_{\psi}
\end{align}
we find
\begin{align}
   \left. \frac{\partial B_{p}}{\partial \psi} \right|_{\theta} &= - \frac{\mu_{0}}{B_{p}} \left. \frac{\partial p}{\partial \psi} \right|_{R} - \frac{I}{R^{2} B_{p}} \frac{d I}{d \psi} - B_{p} \left( \left. \frac{\partial l_{p}}{\partial \theta} \right|_{\psi} \right)^{-1} \left. \frac{\partial}{\partial \psi} \right|_{\theta} \left( \left. \frac{\partial l_{p}}{\partial \theta} \right|_{\psi} \right) \label{eq:BpRadDeriv} \\
   &+ \left( \left. \frac{\partial l_{p}}{\partial \theta} \right|_{\psi} \right)^{-1} \left. \frac{\partial}{\partial \theta} \right|_{\psi} \left[ B_{p} \left( \left. \frac{\partial l_{p}}{\partial \theta} \right|_{\psi} \right)^{-1} \left( \left. \frac{\partial R}{\partial \psi} \right|_{\theta} \left. \frac{\partial R}{\partial \theta} \right|_{\psi} + \left. \frac{\partial Z}{\partial \psi} \right|_{\theta} \left. \frac{\partial Z}{\partial \theta} \right|_{\psi} \right) \right] , \nonumber
\end{align}
where $l_{p}$ is the poloidal arc length defined such that
\begin{align}
   \left. \frac{\partial l_{p}}{\partial \theta} \right|_{\psi} &= \sqrt{\left( \left. \frac{\partial R}{\partial \theta} \right|_{\psi} \right)^{2} + \left( \left. \frac{\partial Z}{\partial \theta} \right|_{\psi} \right)^{2}} . \label{eq:dLpdtheta}
\end{align}
Note that this also allows us to find $d I / d r_{\psi}$ from $dq/dr_{\psi}$ using equation \refEq{eq:safetyFactorDef} after differentiating radially. Lastly we can calculate
\begin{align}
    \vec{\nabla} \alpha &= - \left( I \left. \int_{\theta_{\alpha}}^{\theta} \right|_{\psi} d \theta' \left\{ \frac{1}{R^{2} B_{p}} \left. \frac{\partial l_{p}}{\partial \theta} \right|_{\psi} \left[ \frac{1}{I} \frac{d I}{d \psi} - \frac{1}{B_{p}} \left. \frac{\partial B_{p}}{\partial \psi} \right|_{\theta} - \frac{2}{R} \left. \frac{\partial R}{\partial \psi} \right|_{\theta} \right. \right. \right. \nonumber \\
    &+ \left. \left. \left. \left( \left. \frac{\partial l_{p}}{\partial \theta} \right|_{\psi} \right)^{-1} \left. \frac{\partial}{\partial \psi} \right|_{\theta} \left( \left. \frac{\partial l_{p}}{\partial \theta} \right|_{\psi} \right) \right] \right\} - \left[ \frac{I}{R^{2} B_{p}} \left. \frac{\partial l_{p}}{\partial \theta} \right|_{\psi} \right]_{\theta = \theta_{\alpha}} \frac{d \theta_{\alpha}}{d \psi} + \frac{d \Omega_{\zeta}}{d \psi} t \right) \vec{\nabla} \psi \label{eq:gradAlphaMiller} \\
    &- \frac{I}{R^{2} B_{p}} \left. \frac{\partial l_{p}}{\partial \theta} \right|_{\psi} \vec{\nabla} \theta + \vec{\nabla} \zeta , \nonumber
\end{align}
directly from equation \refEq{eq:alphaDef}, where all quantities are evaluated on the flux surface of interest.

%===================================================%
\subsection{Asymptotic expansion ordering}
\label{subsec:asymptoticExpansion}
%===================================================%

Now we will investigate the effect of high order flux surface shaping on the geometric coefficients and ultimately the fluxes of particles, momentum, and energy. We can always Fourier analyze the flux surface shape (without loss of generality) to write 
\begin{align}
   r_{0} \left( \theta \right) &= r_{\psi 0} \left( 1 - \sum_{m} C_{m} \Cos{m \left( \theta + \theta_{t m} \right)} \right) \label{eq:fluxSurfaceSpec}
\end{align}
and consequently
\begin{align}
   \left. \frac{\partial r}{\partial r_{\psi}} \right|_{r_{\psi 0}, \theta} = 1 - \sum_{m} & \Big[ \left( C_{m} + r_{\psi 0} C'_{m} \right) \Cos{m \left( \theta + \theta_{t m} \right)} \label{eq:fluxSurfaceChangeSpec} \\
   &- m r_{\psi 0} C_{m} \theta'_{t m} \Sin{m \left( \theta + \theta_{t m} \right)} \Big] . \nonumber
\end{align}
Here $m$ is the shaping effect mode number, $C_{m}$ is the mode strength, $\theta_{t m}$ is the poloidal tilt angle of the mode, $C'_{m}$ controls how effectively the mode strength penetrates radially, and $\theta'_{t m}$ indicates how the tilt angle changes radially. The negative sign in front of the Fourier modes in equations \refEq{eq:fluxSurfaceSpec} and \refEq{eq:fluxSurfaceChangeSpec} was chosen so that $m = 2$ with $\theta_{t m} = 0$ corresponds to the traditional vertical elongation, rather than horizontal elongation.

Next we will divide the Fourier modes into low order, indicated by $n$, and high order, indicated by $m$, to get
\begin{align}
   r_{0} \left( \theta \right) &= r_{\psi 0} \left( 1 - \sum_{n} C_{n} \Cos{n \left( \theta + \theta_{t n} \right)} - \sum_{m} C_{m} \Cos{m \left( \theta + \theta_{t m} \right)} \right) \label{eq:fluxSurfaceSpecScaleSep}
\end{align}
and
\begin{align}
   \left. \frac{\partial r}{\partial r_{\psi}} \right|_{r_{\psi 0}, \theta} = 1 &- \sum_{n} \Big[ \left( C_{n} + r_{\psi 0} C'_{n} \right) \Cos{n \left( \theta + \theta_{t n} \right)} - n r_{\psi 0} C_{n} \theta'_{t n} \Sin{n \left( \theta + \theta_{t n} \right)} \Big] \label{eq:fluxSurfaceChangeSpecScaleSep} \\
   &- \sum_{m} \Big[ \left( C_{m} + r_{\psi 0} C'_{m} \right) \Cos{m \left( \theta + \theta_{t m} \right)} - m r_{\psi 0} C_{m} \theta'_{t m} \Sin{m \left( \theta + \theta_{t m} \right)} \Big] . \nonumber
\end{align}
This allows us to order $n \sim 1$ and expand in $m_{c} \gg 1$, where $m_{c}$ is a characteristic high order mode number such that $m \sim m_{c}$ for every $m$. We will use an expansion in $m_{c} \gg 1$ to investigate the effect of high order flux surface shaping on a traditionally shaped equilibrium. The division between $m$ and $n$ is completely general and can be done for any combination of Fourier modes, but we expect that for the expansion to be accurate (and hence useful) there should be a clear separation of scales between the two groups, i.e. $n \ll m_{c} \sim m$.

This expansion distinguishes the long spatial scale coordinate $\theta$, from a short spatial scale coordinate
\begin{align}
  z \equiv m_{c} \theta . \label{eq:zDef}
\end{align}
We can incorporate the short spatial scale by substituting $z$ for $\theta$ in the high order Fourier terms in equations \refEq{eq:fluxSurfaceSpecScaleSep} and \refEq{eq:fluxSurfaceChangeSpecScaleSep} to get $r_{0} \left( \theta, z \right)$ and $\left. \partial r / \partial r_{\psi} \right|_{r_{\psi 0}, \theta, z}$. Furthermore this separation of scales, e.g. $r_{0} \left( \theta, z \right)$, means that
\begin{align}
  \left. \frac{\partial}{\partial \theta} \right|_{w_{||}, \mu} = \left. \frac{\partial}{\partial \theta} \right|_{z, w_{||}, \mu} + m_{c} \left. \frac{\partial}{\partial z} \right|_{\theta, w_{||}, \mu} . \label{eq:poloidalDerivTransform}
\end{align}
Experimentally we are only interested in bulk behavior, so we will eventually average quantities in $z$ using
\begin{align}
  \left\langle \ldots \right\rangle_{z} \equiv \frac{1}{2 \pi} \left. \oint_{-\pi}^{\pi} \right|_{\theta} dz \left( \ldots \right) . \label{eq:zAvg}
\end{align}

%===================================================%
\subsection{Gyrokinetic symmetry}
\label{subsec:gyroSym}
%===================================================%

This section contains an analytic calculation that demonstrates a symmetry of the gyrokinetic model, when expanding in $m_{c} \gg 1$. Since turbulent eddies are generally quite extended along the field line, we expect them to effectively average over the small scale magnetic variations created by high order flux surface shaping. Therefore, we presuppose that tilting such shaping should have a minimal effect on the turbulence. However, the unexpected result of this calculation is that the effect of tilt on the turbulence is exponentially small in $m_{c} \gg 1$, rather than polynomially. Hence we find that tilting high order flux surface shaping has an exponentially small effect on the turbulent fluxes. This argument only relies on $m_{c} \gg 1$ and does \textit{not} presume that the flux surface shaping is weak.

We will start with a completely general local equilibrium, with flux surfaces specified by $r_{0} \left( \theta, z \left( \theta \right) \right)$ and $\left. \partial r / \partial r_{\psi} \right|_{r_{\psi 0}, \theta, z \left( \theta \right)}$ (see equations \refEq{eq:fluxSurfaceSpecScaleSep} through \refEq{eq:zDef}). Using this specification we will compare two different geometries that are identical except for the form of $z \left( \theta \right)$. In the untilted case $z \left( \theta \right) = z_{u} \left( \theta \right) \equiv m_{c} \theta$, while in the tilted case $z \left( \theta \right) = z_{t} \left( \theta \right) \equiv m_{c} \left( \theta + \theta_{t} \right)$. The $n \sim 1$ shaping effects are not tilted. We see that the tilted case includes a single global tilt of all the high order shaping effects (those that scale with $m_{c}$). This alters the equilibrium and in principle changes the transport properties, but we will show its effect is exponentially small when expanding in $m_{c} \gg 1$.

Although we just presented two specific examples of $z \left( \theta \right)$, we are free to calculate the geometric coefficients for a completely general $z \left( \theta \right)$. From the form of the ten geometric coefficients (see equations \refEq{eq:gyrokineticEq} through \refEq{eq:parallelAcceleration}) we see that $z \left( \theta \right) $ only enters as $z$, derivatives of $z$, and in the integral over poloidal angle contained in $\vec{\nabla} \alpha$ (see equation \refEq{eq:gradAlphaMiller}). This means that we can indicate the poloidal dependence of any geometric coefficient, $Q_{\text{geo}} = \left\{ B, \hat{b} \cdot \vec{\nabla} \theta, v_{d s \psi}, v_{d s \alpha}, a_{s ||}, \left| \vec{\nabla} \psi \right|^{2}, \vec{\nabla} \psi \cdot \vec{\nabla} \alpha, \left| \vec{\nabla} \alpha \right|^{2}, R, \left. \partial R / \partial \psi \right|_{\theta} \right\}$, by writing it as
\begin{align}
   Q_{\text{geo}} \left( \theta, z, \frac{\partial z}{\partial \theta}, \frac{\partial^{2} z}{\partial \theta^{2}}, \left. \int_{\theta_{\alpha}}^{\theta} \right|_{\psi} d \theta' F_{\alpha} \left( \theta', z \left( \theta' \right), \frac{\partial z}{\partial \theta'}, \frac{\partial^{2} z}{\partial \theta'^{2}} \right) - \left[ \frac{1}{R^{2} B_{p}^{2}} \left. \frac{\partial l_{p}}{\partial \theta} \right|_{\psi} \right]_{\theta = \theta_{\alpha}} \frac{d \theta_{\alpha}}{d \psi} \right) , \label{eq:Qgeo}
\end{align}
where
\begin{align}
   F_{\alpha} \left( \theta, z \left( \theta \right), \frac{\partial z}{\partial \theta}, \frac{\partial^{2} z}{\partial \theta^{2}} \right) &\equiv \frac{1}{R^{2} B_{p}} \left. \frac{\partial l_{p}}{\partial \theta} \right|_{\psi} \Bigg[ \frac{1}{I} \frac{d I}{d \psi} - \frac{1}{B_{p}} \left. \frac{\partial B_{p}}{\partial \psi} \right|_{\theta} - \frac{2}{R} \left. \frac{\partial R}{\partial \psi} \right|_{\theta} \nonumber \\
    &+ \left( \left. \frac{\partial l_{p}}{\partial \theta} \right|_{\psi} \right)^{-1} \left. \frac{\partial}{\partial \psi} \right|_{\theta} \left( \left. \frac{\partial l_{p}}{\partial \theta} \right|_{\psi} \right) \Bigg] \label{eq:FalphaDef}
\end{align}
is a periodic function of both $\theta$ and $z$.

Now we will compare the untilted equilibrium ($z \left( \theta \right) = z_{u} \left( \theta \right) \equiv m_{c} \theta$) and the equilibrium with tilted high order shaping effects ($z \left( \theta \right) = z_{t} \left( \theta \right) \equiv m_{c} \left( \theta + \theta_{t} \right)$). Since the only difference between the two cases is contained in the form of $z \left( \theta \right)$, we only need to look for differences in the arguments of equation \refEq{eq:Qgeo}. We immediately see that $\partial z_{u} / \partial \theta = \partial z_{t} / \partial \theta = m_{c}$ and $\partial^{2} z_{u} / \partial \theta^{2} = \partial^{2} z_{t} / \partial \theta^{2} = 0$, so we can eliminate the derivates to write the geometric coefficients as
\begin{align}
   Q_{\text{geo}} \left( \theta, z, G_{\alpha}^{\theta} \left( \theta, z \left( \theta \right) \right) + H_{\alpha} \right) \label{eq:QgeoSimple}
\end{align}
for both cases, where we choose to define
\begin{align}
   G_{\alpha}^{\theta} \left( \theta, z \left( \theta \right) \right) \equiv& \left. \int_{\theta_{\alpha}}^{\theta} \right|_{\psi} d \theta' F_{\alpha} \left( \theta', z \left( \theta' \right) \right) \label{eq:GalphaThetaDef} \\
   H_{\alpha} \equiv& - \left[ \frac{1}{R^{2} B_{p}^{2}} \left. \frac{\partial l_{p}}{\partial \theta} \right|_{\psi} \right]_{\theta = \theta_{\alpha}} \frac{d \theta_{\alpha}}{d \psi} .
\end{align}

As we will now show, we can also eliminate the integral $\left. \int_{\theta_{\alpha}}^{\theta} \right|_{\psi} d \theta' F_{\alpha} \left( \theta', z \left( \theta' \right) \right)$, in addition to the derivatives. An alternative method to do this is given in \ref{app:alphaIntegral}, but here we will start by defining the operator
\begin{align}
   \Lambda \left[ g \right] \left( \theta, z \right) &\equiv \left. \int_{z_{0}}^{z} \right|_{\theta} d z' \Big( g \left( \theta, z' \right) - \left\langle g \left( \theta, z \right) \right\rangle_{z} \Big) , \label{eq:Idef}
\end{align}
where the integral over $z$ is done holding $\theta$ constant, $g \left( \theta, z \right)$ is a yet unspecified function that is periodic in both $\theta$ and $z$, and $z_{0}$ is chosen such that $\left\langle \Lambda \left[ g \right] \left( \theta, z \right) \right\rangle_{z} = 0$ (which can always be found when $g$ is periodic in $z$). Taking the total derivative in $\theta$ we find
\begin{align}
   \frac{d}{d \theta} \Lambda \left[ g \right] \left( \theta, z \left( \theta \right) \right) &= \left. \frac{\partial}{\partial \theta} \right|_{z} \Lambda \left[ g \right] + m_{c} \left. \frac{\partial}{\partial z} \right|_{\theta} \Lambda \left[ g \right] ,
\end{align}
where we have taken $d z / d \theta = m_{c}$. Substituting in equation \refEq{eq:Idef} and rearranging gives
\begin{align}
   g \left( \theta, z \right) - \left\langle g \left( \theta, z \right) \right\rangle_{z} &= \frac{1}{m_{c}} \frac{d}{d \theta} \Lambda \left[ g \right] - \frac{1}{m_{c}} \left. \frac{\partial}{\partial \theta} \right|_{z} \Lambda \left[ g \right] . \label{eq:gDiff}
\end{align}
Taking the integral of equation \refEq{eq:gDiff} and then using $\left. \partial \left( \Lambda \left[ g \right] \right) / \partial \theta \right|_{z} = \Lambda \left[ \left. \partial g/ \partial \theta \right|_{z} \right]$ gives
\begin{align}
   \left. \int_{\theta_{\alpha}}^{\theta} \right|_{\psi} d \theta' \Big( g \left( \theta', z \left( \theta' \right) \right) - \left\langle g \left( \theta', z \right) \right\rangle_{z} \Big) &= \frac{1}{m_{c}} \Big( \Lambda \left[ g \right] \left( \theta, z \right) - \Lambda \left[ g \right] \left( \theta_{\alpha}, z \left( \theta_{\alpha} \right) \right) \Big) \label{eq:recursionRel} \\
   &- \frac{1}{m_{c}} \left. \int_{\theta_{\alpha}}^{\theta} \right|_{\psi} d \theta' \Lambda \left[ \left. \frac{\partial g}{\partial \theta'} \right|_{z} \right] . \nonumber
\end{align}
Now if we make the substitution $g \left( \theta, z \right) \rightarrow \Lambda \left[ \left. \partial g/ \partial \theta \right|_{z} \right]$ in equation \refEq{eq:recursionRel}, we obtain
\begin{align}
   \left. \int_{\theta_{\alpha}}^{\theta} \right|_{\psi} d \theta' \Lambda \left[ \left. \frac{\partial g}{\partial \theta'} \right|_{z} \right] &= \frac{1}{m_{c}} \left( \Lambda^{2} \left[ \left. \frac{\partial g}{\partial \theta} \right|_{z} \right] \left( \theta, z \left( \theta \right) \right) - \Lambda^{2} \left[ \left. \frac{\partial g}{\partial \theta} \right|_{z} \right]  \left( \theta_{\alpha}, z \left( \theta_{\alpha} \right) \right) \right) \label{eq:recursionRelSecondIteration} \\
   &- \frac{1}{m_{c}} \left. \int_{\theta_{\alpha}}^{\theta} \right|_{\psi} d \theta' \Lambda^{2} \left[ \left. \frac{\partial^2 g}{\partial \theta'^2} \right|_{z} \right] \nonumber
\end{align}
because, since $\left\langle \Lambda \left[ g \right] \right\rangle_{z} = 0$, we know that $\left. \partial \left\langle \Lambda \left[ g \right] \right\rangle_{z} / \partial \theta \right|_{z} = \left\langle \Lambda \left[ \left. \partial g/ \partial \theta \right|_{z} \right] \right\rangle_{z} = 0$. Here $\Lambda^{i} \left[ \ldots \right]$ indicates that the operator $\Lambda \left[ \ldots \right]$ is applied $i$ times. Substituting equation \refEq{eq:recursionRelSecondIteration} into the last term of equation \refEq{eq:recursionRel}, we see that equation \refEq{eq:recursionRel} is a recursion relation that can be put in the form of an infinite series,
\begin{align}
   \left. \int_{\theta_{\alpha}}^{\theta} \right|_{\psi} d \theta' \Big( g \left( \theta', z \left( \theta' \right) \right) - \left\langle g \left( \theta', z \right) \right\rangle_{z} \Big) &= \sum_{p = 1}^{\infty} \frac{\left( - 1 \right)^{p-1}}{m_{c}^{p}} \left( \Lambda^{p} \left[ \left. \frac{\partial^{p-1} g}{\partial \theta^{p-1}} \right|_{z} \right] \left( \theta, z \left( \theta \right) \right) \right. \\
   &- \left. \Lambda^{p} \left[ \left. \frac{\partial^{p-1} g}{\partial \theta^{p-1}} \right|_{z} \right]  \left( \theta_{\alpha}, z \left( \theta_{\alpha} \right) \right) \right) . \nonumber
\end{align}
Finally by substituting $g \left( \theta, z \right) = F_{\alpha} \left( \theta, z \right)$ and rearranging we can calculate the integral appearing in the geometric coefficients (see equation \refEq{eq:QgeoSimple}) to be
\begin{align}
   G_{\alpha}^{\theta} \left( \theta, z \left( \theta \right) \right) &= \left.\int_{\theta_{\alpha}}^{\theta} \right|_{\psi} d \theta' \left\langle F_{\alpha} \left( \theta', z \right) \right\rangle_{z} + \sum_{p = 1}^{\infty} \frac{\left( - 1 \right)^{p-1}}{m_{c}^{p}} \left( \Lambda^{p} \left[ \left. \frac{\partial^{p-1} F_{\alpha}}{\partial \theta^{p-1}} \right|_{z} \right] \left( \theta, z \left( \theta \right) \right) \right. \nonumber \\
   &- \left. \Lambda^{p} \left[ \left. \frac{\partial^{p-1} F_{\alpha}}{\partial \theta^{p-1}} \right|_{z} \right] \left( \theta_{\alpha}, z \left( \theta_{\alpha} \right) \right) \right) . \label{eq:generalAlphaInt}
\end{align}

For the untilted case we choose $\theta_{\alpha} = d \theta_{\alpha} / d \psi = 0$ such that the quantity appearing in the geometric coefficients (see equation \refEq{eq:QgeoSimple}) becomes
\begin{align}
   G_{\alpha}^{\theta} \left( \theta, z_{u} \left( \theta \right) \right) &+ H_{\alpha} = \left. \int_{0}^{\theta} \right|_{\psi} d \theta' F_{\alpha} \left( \theta', z_{u} \left( \theta' \right) \right) .
\end{align}
Hence substituting equation \refEq{eq:generalAlphaInt} (remembering that $\theta_{\alpha} = 0$) gives
\begin{align}
   G_{\alpha}^{\theta} \left( \theta, z_{u} \left( \theta \right) \right) &+ H_{\alpha} = \left. \int_{0}^{\theta} \right|_{\psi} d \theta' \left\langle F_{\alpha} \left( \theta', z \right) \right\rangle_{z} \label{eq:untiltAlphaInt} \\
   &+ \sum_{p = 1}^{\infty} \frac{\left( - 1 \right)^{p-1}}{m_{c}^{p}} \left( \Lambda^{p} \left[ \left. \frac{\partial^{p-1} F_{\alpha}}{\partial \theta^{p-1}} \right|_{z} \right] \left( \theta, z_{u} \left( \theta \right) \right) - \Lambda^{p} \left[ \left. \frac{\partial^{p-1} F_{\alpha}}{\partial \theta^{p-1}} \right|_{z} \right] \left( 0, 0 \right) \right) . \nonumber
\end{align}
In the tilted case ($z = z_{t} = m_{c} \left( \theta + \theta_{t} \right)$) we can carefully choose
\begin{align}
   \frac{d \theta_{\alpha}}{d \psi} &= \left[ \frac{1}{R^{2} B_{p}^{2}} \left. \frac{\partial l_{p}}{\partial \theta} \right|_{\psi} \right]_{\theta = \theta_{\alpha}}^{-1} \left\{ \left. \int_{\theta_{\alpha}}^{0} \right|_{\psi} d \theta' \left\langle F_{\alpha} \left( \theta', z \right) \right\rangle_{z} \right. \label{eq:thetaAlphaDerivTilted} \\
   &+ \left. \sum_{p = 1}^{\infty} \frac{\left( - 1 \right)^{p-1}}{m_{c}^{p}} \left( \Lambda^{p} \left[ \left. \frac{\partial^{p-1} F_{\alpha}}{\partial \theta^{p-1}} \right|_{z} \right] \left( 0, 0 \right) - \Lambda^{p} \left[ \left. \frac{\partial^{p-1} F_{\alpha}}{\partial \theta^{p-1}} \right|_{z} \right] \left( \theta_{\alpha}, z_{t} \left( \theta_{\alpha} \right) \right) \right) \right\} \nonumber
\end{align}
to get 
\begin{align}
   G_{\alpha}^{\theta} \left( \theta, z_{t} \left( \theta \right) \right) &+ H_{\alpha} = \left. \int_{0}^{\theta} \right|_{\psi} d \theta' \left\langle F_{\alpha} \left( \theta', z \right) \right\rangle_{z} \label{eq:tiltAlphaInt} \\
   &+ \sum_{p = 1}^{\infty} \frac{\left( - 1 \right)^{p-1}}{m_{c}^{p}} \left( \Lambda^{p} \left[ \left. \frac{\partial^{p-1} F_{\alpha}}{\partial \theta^{p-1}} \right|_{z} \right] \left( \theta, z_{t} \left( \theta \right) \right) - \Lambda^{p} \left[ \left. \frac{\partial^{p-1} F_{\alpha}}{\partial \theta^{p-1}} \right|_{z} \right] \left( 0, 0 \right) \right) , \nonumber
\end{align}
which exactly matches equation \refEq{eq:untiltAlphaInt} (except for replacing $z_{u}$ with $z_{t}$). This means the entire effect of the tilt can be contained in the functional form of $z$. To make things as simple as possible we also choose
\begin{align}
   \theta_{\alpha} = 0 \label{eq:thetaAlphaTilted}
\end{align}
for the tilted case.

This means that the geometric coefficients for both the untilted and tilted cases can be written in the form
\begin{align}
   Q_{\text{geo}} \left( \theta, z \right) , \label{eq:QgeoFinal}
\end{align}
where $z = z_{u}$ for the untilted case and $z = z_{t}$ for the tilted case. Therefore, we know that
\begin{align}
   Q_{\text{geo}}^{t} \left( \theta, z_{u} \right) = Q_{\text{geo}}^{u} \left( \theta, z_{u} + m_{c} \theta_{t} \right)
\end{align}
for each of the geometric coefficients.

Since all of the geometric coefficients depend separately on both $\theta$ and $z$ we know that $h_{s}$, $\phi$, $A_{||}$, and $B_{||}$ must also. The only other way the poloidal coordinate enters the gyrokinetic equations is through the poloidal derivative in the streaming term, but equation \refEq{eq:poloidalDerivTransform} is appropriate for both $z = z_{u}$ and $z = z_{t}$. Hence, using any solution to the gyrokinetic equation for the untilted case, $\left\{ h_{s}^{u} \left( \theta, z_{u} \right), \phi^{u} \left( \theta, z_{u} \right), A_{||}^{u} \left( \theta, z_{u} \right), B_{||}^{u} \left( \theta, z_{u} \right) \right\}$, we can construct a solution for the tilted case,
\begin{align}
  &\left\{ h_{s}^{t} \left( \theta, z_{u} \right), \phi^{t} \left( \theta, z_{u} \right), A_{||}^{t} \left( \theta, z_{u} \right), B_{||}^{t} \left( \theta, z_{u} \right) \right\} \label{eq:symSolToAsymSol} \\ 
  &\hspace{3em} = \left\{ h_{s}^{u} \left( \theta, z_{u} + m_{c} \theta_{t} \right), \phi^{u} \left( \theta, z_{u} + m_{c} \theta_{t} \right), A_{||}^{u} \left( \theta, z_{u} + m_{c} \theta_{t} \right), B_{||}^{u} \left( \theta, z_{u} + m_{c} \theta_{t} \right) \right\} , \nonumber
\end{align}
given our choices for the free parameter $\theta_{\alpha} \left( \psi \right)$ in the definition of $\alpha$ (see equations \refEq{eq:thetaAlphaDerivTilted} and \refEq{eq:thetaAlphaTilted}). Because the average over $z$ (see equation \refEq{eq:zAvg}) can always be shifted by $m_{c} \theta_{t}$ without affecting the result these two solution sets give the same large scale turbulent fluxes and turbulent energy exchange between species, e.g. in the electrostatic limit they are
\begin{align}
  \Gamma_{s}^{\phi} &= \frac{4 \pi^{2} i}{m_{s} V'} \left\langle \sum_{k_{\psi}, k_{\alpha}} k_{\alpha} \oint d \theta \left\langle J B \phi \left( k_{\psi}, k_{\alpha} \right) \int dw_{||} d \mu ~ h_{s} \left( - k_{\psi}, - k_{\alpha} \right) J_{0} \left( k_{\perp} \rho_{s} \right) \right\rangle_{z} \right\rangle_{\Delta t} \label{eq:partFluxAvg} \\
  \Pi_{s}^{\phi} &= \frac{4 \pi^{2} i}{V'} \left\langle \sum_{k_{\psi}, k_{\alpha}} k_{\alpha} \oint d \theta \left\langle J B \phi \left( k_{\psi}, k_{\alpha} \right) \int dw_{||} d \mu ~ h_{s} \left( - k_{\psi}, - k_{\alpha} \right) \right. \right. \label{eq:momFluxAvg} \\
   &\times \left. \left. \left[ \left( \frac{I}{B} w_{||} + R^{2} \Omega_{\zeta} \right) J_{0} \left( k_{\perp} \rho_{s} \right) + \frac{i}{\Omega_{s}} \frac{k^{\psi}}{B} \frac{\mu B}{m_{s}} \frac{2 J_{1} \left( k_{\perp} \rho_{s} \right)}{k_{\perp} \rho_{s}} \right] \right\rangle_{z} \right\rangle_{\Delta t} \nonumber \\
  Q_{s}^{\phi} &= \frac{4 \pi^{2} i}{V'} \left\langle \sum_{k_{\psi}, k_{\alpha}} k_{\alpha} \oint d \theta \left\langle J B \phi \left( k_{\psi}, k_{\alpha} \right) \int dw_{||} d \mu ~ h_{s} \left( - k_{\psi}, - k_{\alpha} \right) \right. \right. \label{eq:heatFluxAvg} \\
   &\times \left. \left. \left( \frac{w^{2}}{2} + \frac{Z_{s} e \Phi_{0}}{m_{s}} - \frac{1}{2} R^{2} \Omega_{\zeta}^{2} \right) J_{0} \left( k_{\perp} \rho_{s} \right) \right\rangle_{z} \right\rangle_{\Delta t} \nonumber \\
   P_{Q s}^{\phi} &= \frac{4 \pi^{2}}{V'} \left\langle \sum_{k_{\psi}, k_{\alpha}} \oint d \theta \left\langle J \Omega_{s} \frac{\partial}{\partial t} \left( \phi \left( k_{\psi}, k_{\alpha} \right) \right) \int dw_{||} d \mu ~ h_{s} \left( - k_{\psi}, - k_{\alpha} \right) J_{0} \left( k_{\perp} \rho_{s} \right) \right\rangle_{z} \right\rangle_{\Delta t} . \label{eq:heatingAvg}
\end{align}
Looking at the full electromagnetic fluxes and the turbulent energy exchange between species (see \ref{app:fluxes}) we see that they also remain unchanged by the tilt.

Since we relied on expanding in $m_{c} \gg 1$ to separate scales in equations \refEq{eq:partFluxAvg} through \refEq{eq:heatFluxAvg}, this argument can only give the fluxes as an expansion in powers of $1 / m_{c}$, not the unexpanded quantity. We already know that, since the two configuration are not \textit{exactly} identical, they will in general produce different fluxes. However, the above argument proves the two configurations must have the same fluxes to all orders in $1 / m_{c}$. This demonstrates that, while the fluxes from the two configurations can be different, the difference does not scale polynomially and so cannot scale more strongly than $\sim \Exp{- \beta m_{c}^{\gamma}}$, where $\beta$ and $\gamma$ are both positive and do not depend on $m_{c}$.

%===================================================%
\subsection{Accuracy of the local equilibrium approximation}
\label{subsec:localEqApprox}
%===================================================%

We finish with an important remark concerning our use of an approximate local MHD equilibrium, as opposed to the full global MHD equilibria. Although there was no problem in the Miller local equilibrium, it may not be possible to exactly tilt the high order flux surface shaping poloidally in a real global equilibrium. We can always prescribe a flux surface shape and Fourier analyze it and its radial derivative (see equations \refEq{eq:fluxSurfaceSpec} and \refEq{eq:fluxSurfaceChangeSpec}). We can also use the external shaping coils to arbitrarily tilt the fast shaping of the flux surface of interest. However, the way that the radial derivative of the flux surface shape changes with tilt is set by the global MHD equilibrium and is not under our control (as it is in the Miller local equilibrium approximation). The global equilibrium in a screw pinch has cylindrical symmetry, but in a tokamak toroidal effects preclude tilting the radial derivative of the flux surface shape in the same manner we tilted the flux surface shape itself. 

This means that, strictly speaking, when we introduce $z_{t} \left( \theta \right) = m_{c} \left( \theta + \theta_{t} \right)$ into the derivative of the flux surface shape we are no longer modeling a physically possible tokamak. However, we can show that this error is exponentially small by rearranging the Grad-Shafranov equation as
\begin{align}
   \nabla^{2} \psi + I \frac{d I}{d \psi} = \frac{2}{R} \vec{\nabla} R \cdot \vec{\nabla} \psi - \mu_{0} R^{2} \left. \frac{\partial p}{\partial \psi} \right|_{R} . \label{eq:gradShafranovMod}
\end{align}
The left side of this equation is completely cylindrically symmetric, while the right side contains all of the toroidal effects, which only enter through $R \left( r_{\psi}, \theta \right)$ (see equation \refEq{eq:geoMajorRadius}). Taylor expanding equation \refEq{eq:gradShafranovMod} in $\epsilon \equiv r_{\psi 0} / R_{0} \ll 1$ (the inverse aspect ratio), we see that toroidicity affects different shaping modes at different orders. For example, reference \cite{HakkarainenEquilibrium1990} shows that at $O \left( 1 \right)$ the equation is entirely cylindrically symmetric, at $O \left( \epsilon \right)$ toroidicity introduces a natural shift (i.e. the Shafranov shift), at $O \left( \epsilon^{2} \right)$ toroidicity introduces a natural elongation, and at $O \left( \epsilon^{3} \right)$ toroidicity introduces a natural triangularity. This indicates that, in a global equilibrium, toroidicity introduces an $O \left( \epsilon^{m} \right)$ modification to the $m^{\text{th}}$ Fourier mode of a flux surface. Therefore, the error introduced into the geometric coefficients by ignoring this effect in the local equilibrium approximation is $O \left( \epsilon^{m_{\text{min}}} \right)$, where $m_{\text{min}}$ is the smallest mode number that is tilted. This error is exponentially small in $m_{\text{min}} \gg 1$, hence it does not change our result that tilting the equilibrium has an exponentially small effect on the turbulent fluxes.

%===================================================%
\section{Numerical results}
\label{sec:numResults}
%===================================================%

In this section we will give numerical results to test the analytic conclusions of the previous section. We use GS2 \cite{DorlandETGturb2000}, a local $\delta f$ gyrokinetic code, to calculate the nonlinear turbulent fluxes of particles, momentum, and energy generated by a given geometry. We investigate the influence of the shape of the flux surface of interest by scanning $m_{c}$, the mode number of the poloidal shaping effect. The geometry is specified using the generalization of the Miller local equilibrium model presented in section \ref{subsec:millerGeo}. The flux surface shapes (shown in figure \ref{fig:simGeo}) are parameterized by equations \refEq{eq:fluxSurfaceSpecScaleSep} and \refEq{eq:fluxSurfaceChangeSpecScaleSep}, with only one high order mode, $m = m_{c}$, and no low order modes (i.e. $C_{n} = 0$). We will choose $C_{m} = 3 / \left( 4 m_{c}^{2} \right)$, $C_{m}' = \left( m_{c} - 2 \right) C_{m} / r_{\psi 0}$, and $\theta_{t m}' = 0$ in the scan to give the flux surfaces a reasonable shape. Up-down asymmetric geometries are created by fixing the tilt angle at $\theta_{t m} = \pi / \left( 2 m_{c} \right)$, the angle halfway between neighboring up-down symmetric configurations (at $\theta_{t m} = 0$ and $\theta_{t m} = \pi / m_{c}$).

Except for the flux surface shape, all parameters are fixed at Cyclone base case values \cite{DimitsCycloneBaseCase2000}: a minor radius of $r_{\psi 0} / a= 0.54$, a major radius of $R_{0} / a = 3$, a safety factor of $q = 1.4$, a magnetic shear of $\hat{s} = 0.8$, a temperature gradient of $a / L_{T s} = 2.3$, and a density gradient of $a / L_{n s} = 0.733$, where $a$ is the tokamak minor radius. All simulations are electrostatic and collisionless. The fluxes calculated by GS2 are normalized to their gyroBohm value, which for the momentum flux is $\Pi_{gB} = \rho_{th i}^{2} n_{i} m_{i} v_{th i}^{2} / a$, where $\rho_{th i} = v_{th i} / \Omega_{i}$ is the ion thermal gyroradius and $v_{th i} = \sqrt{2 T_{i} / m_{i}}$ is the ion thermal velocity.

We will compare the numerical scans in $m_{c}$ (shown in figure \ref{fig:simGeo}) to the analytic theory in two different manners. From equation \refEq{eq:symSolToAsymSol} we expect that, given the poloidal distribution of any flux from a geometry with high order shaping, it should be possible to predict the flux from any geometry that is identical except for a poloidal tilt of the high order shaping. First, we will directly investigate this by comparing the poloidal dependence of the fluxes of particles, momentum, and energy in just such geometries.  Then, we will show that the change in the total fluxes due to tilting high order shaping disappears in the limit of $m_{c} \gg 1$.

%===================================================%
\subsection{Poloidal structure of fluxes}
\label{subsec:polStructComp}
%===================================================%

\begin{figure}
 \centering

 \includegraphics[width=0.18\textwidth]{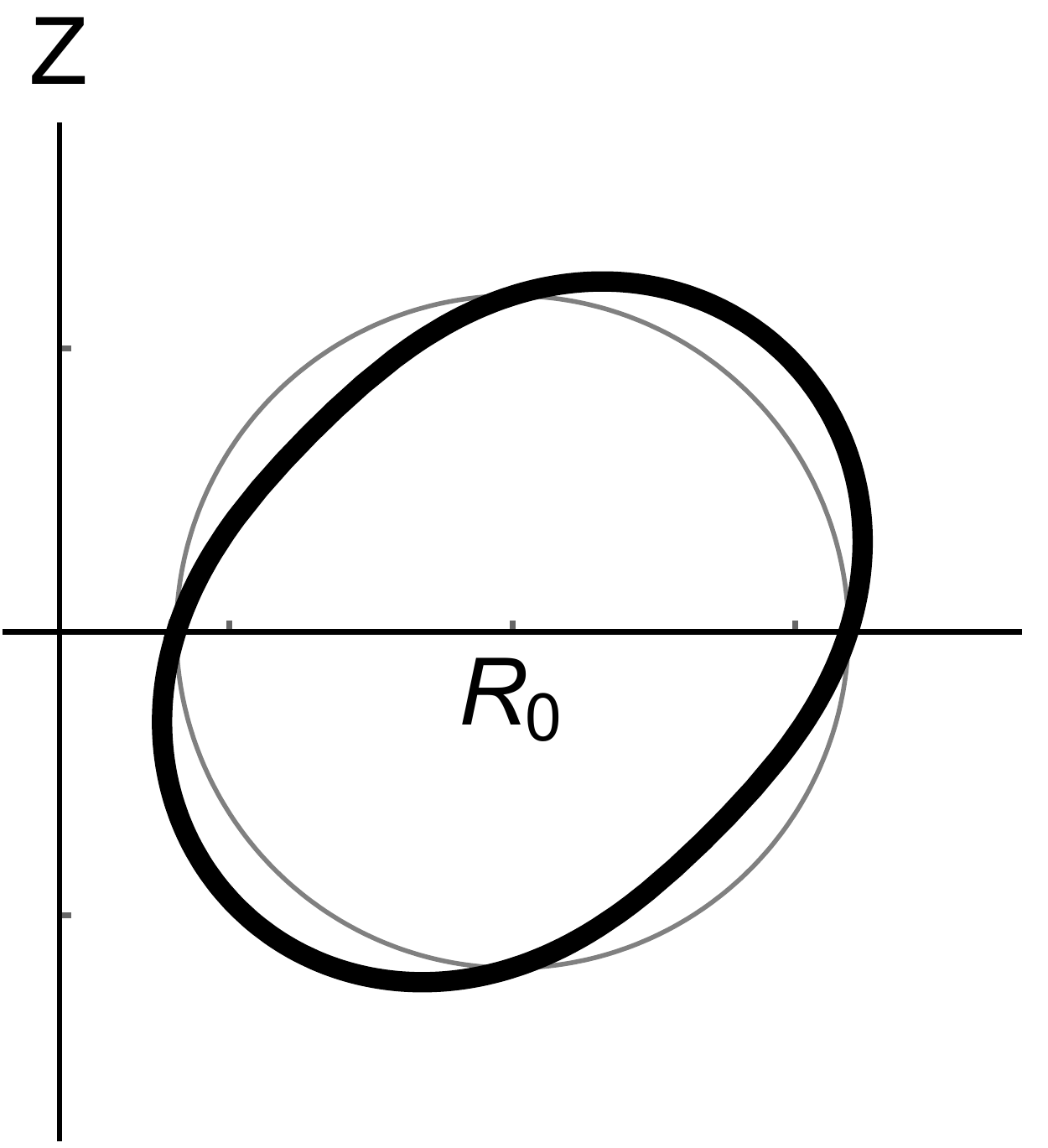}
 \includegraphics[width=0.18\textwidth]{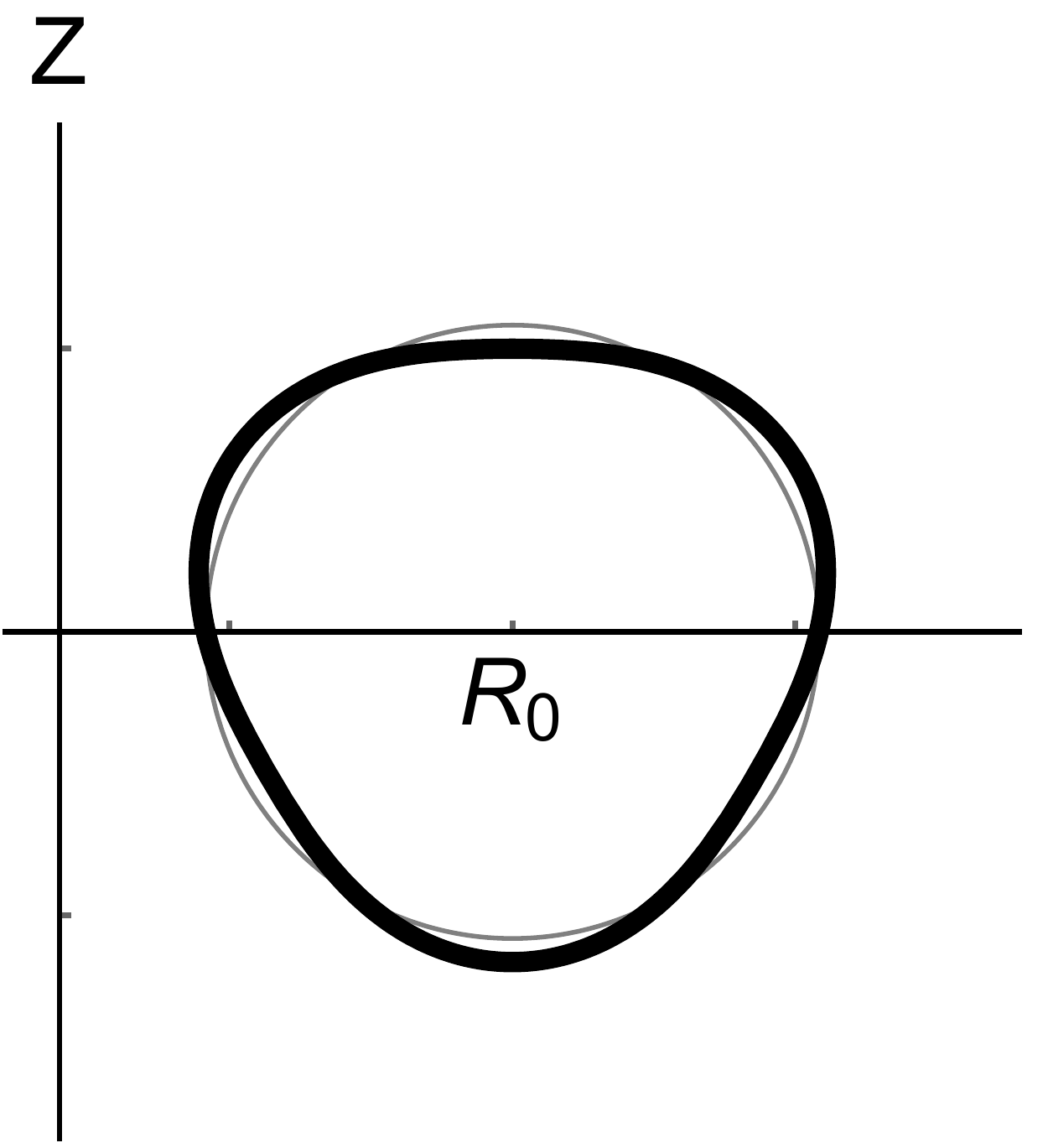}
 \includegraphics[width=0.18\textwidth]{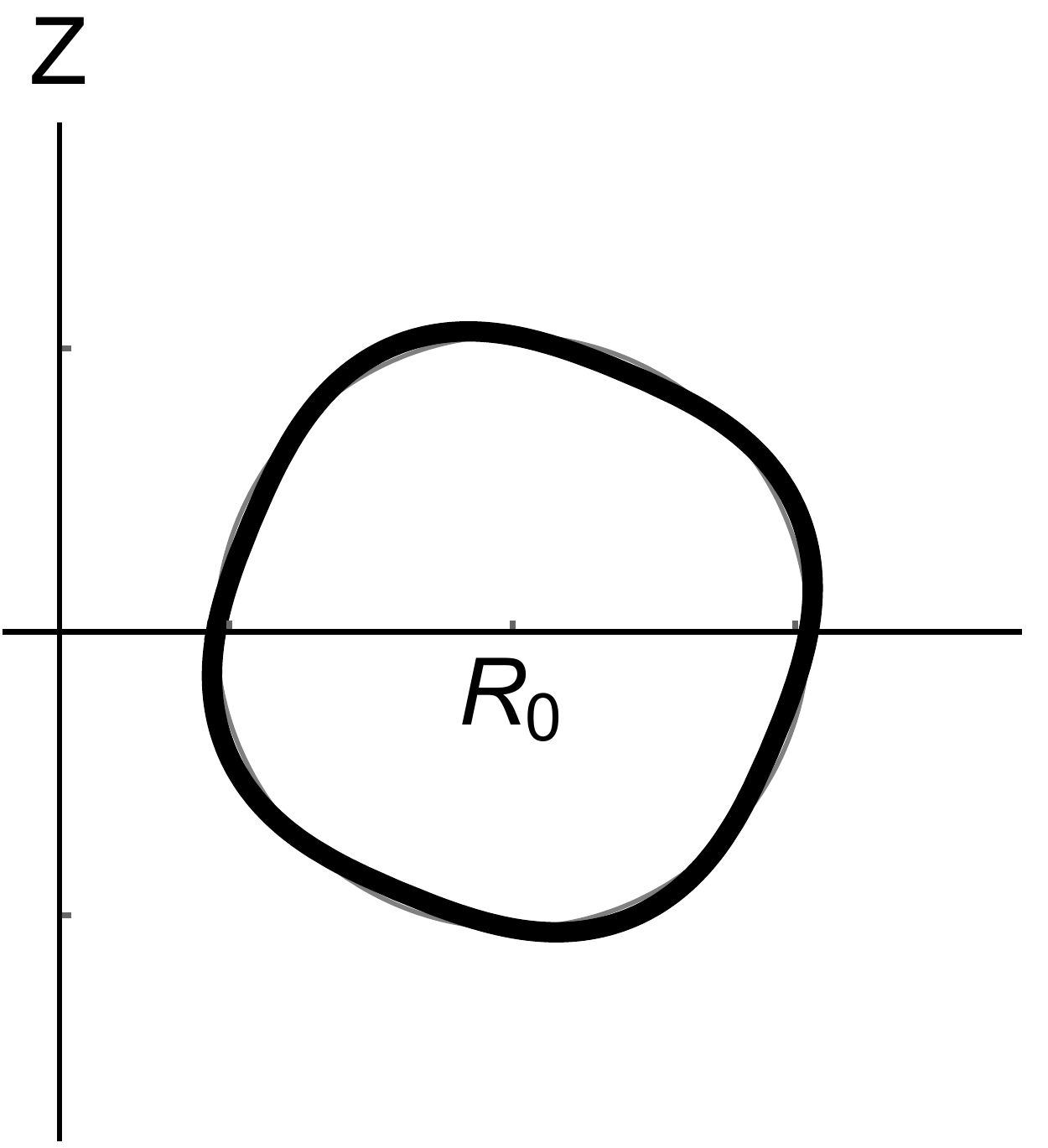}
 \includegraphics[width=0.18\textwidth]{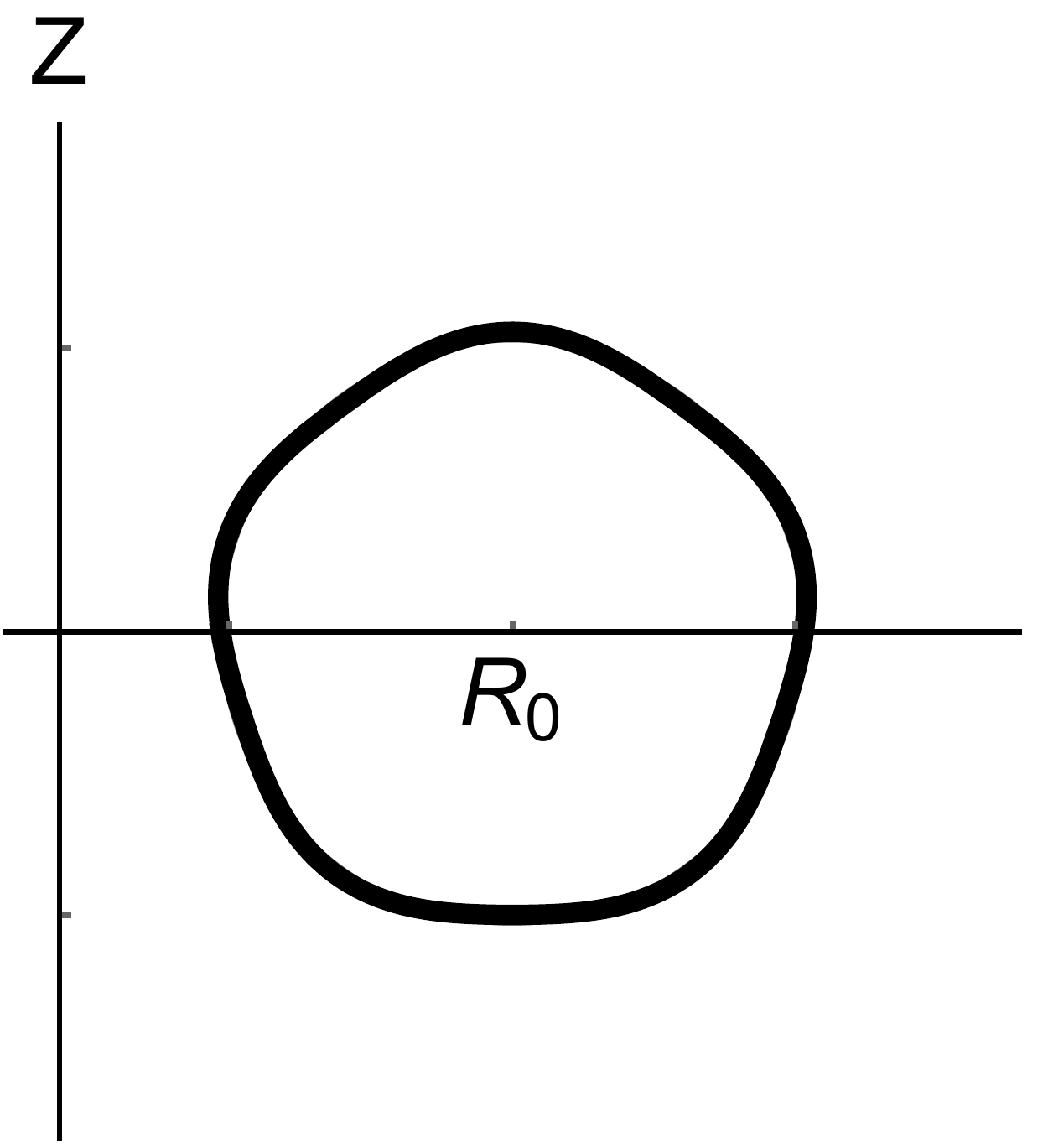}
 \includegraphics[width=0.18\textwidth]{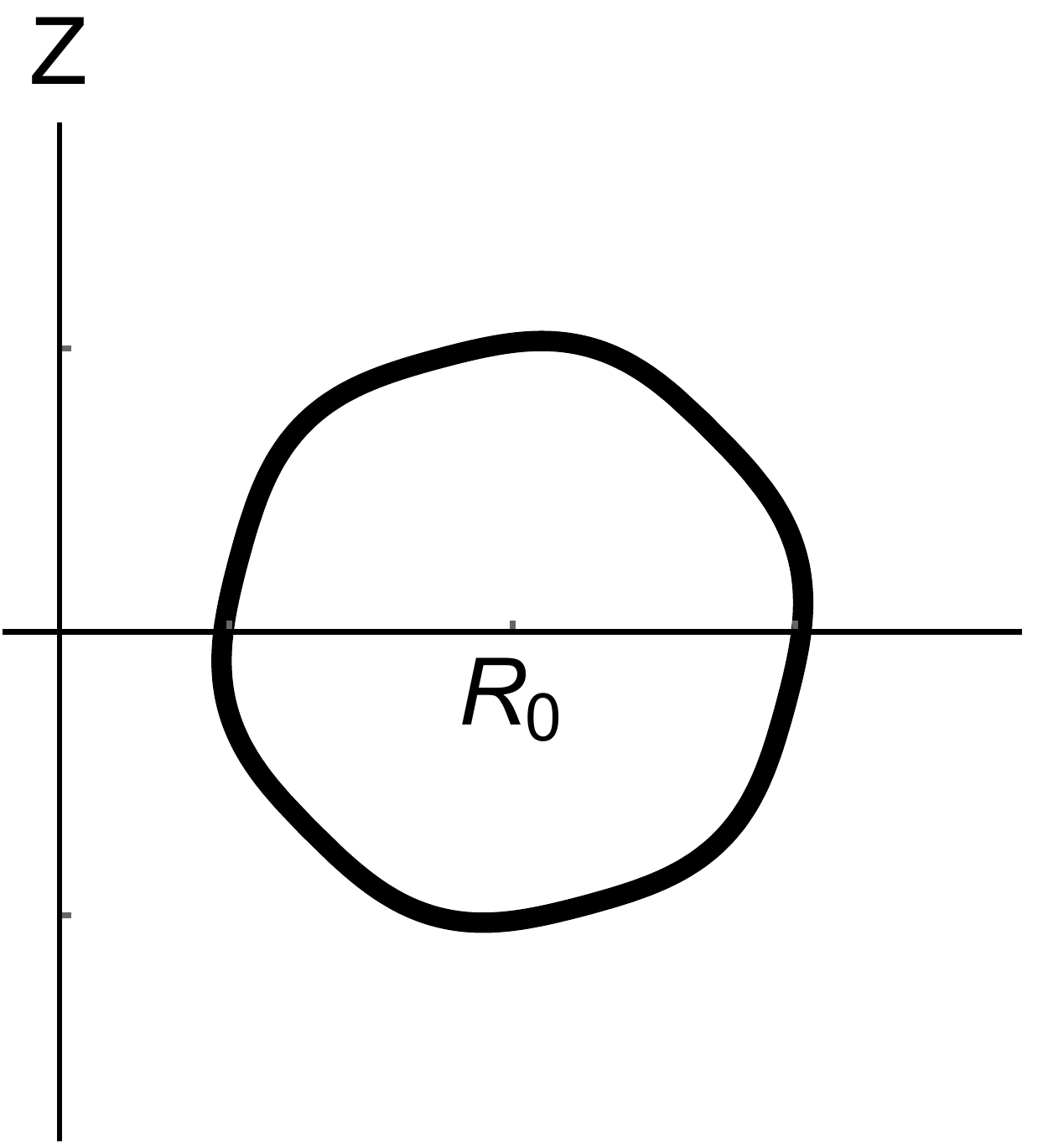}
 \includegraphics[width=0.0111\textwidth]{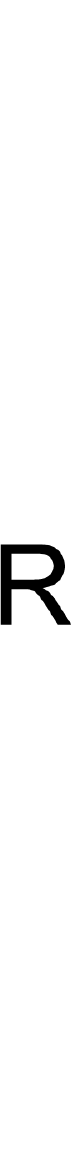}

 \includegraphics[width=0.18\textwidth]{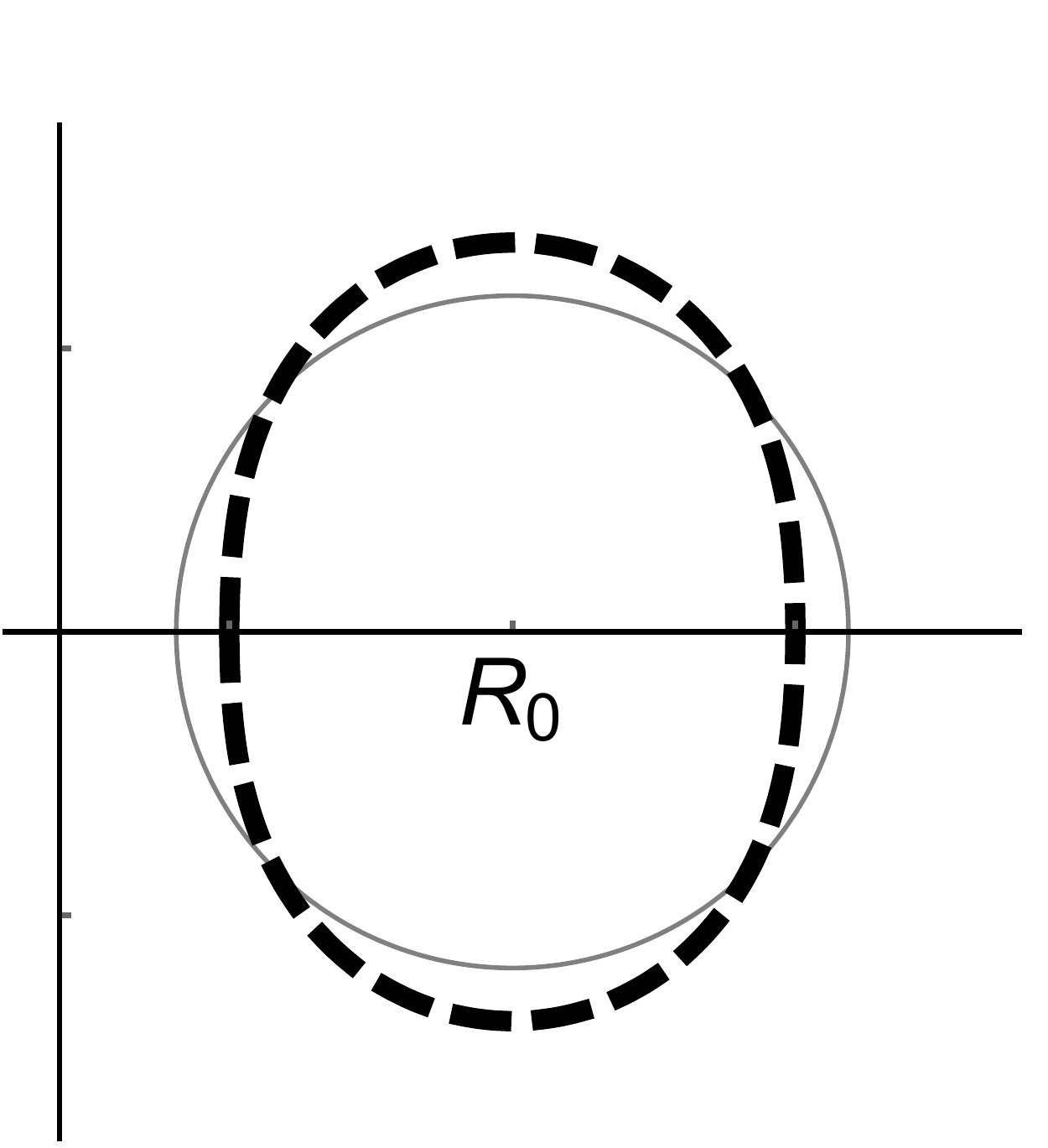}
 \includegraphics[width=0.18\textwidth]{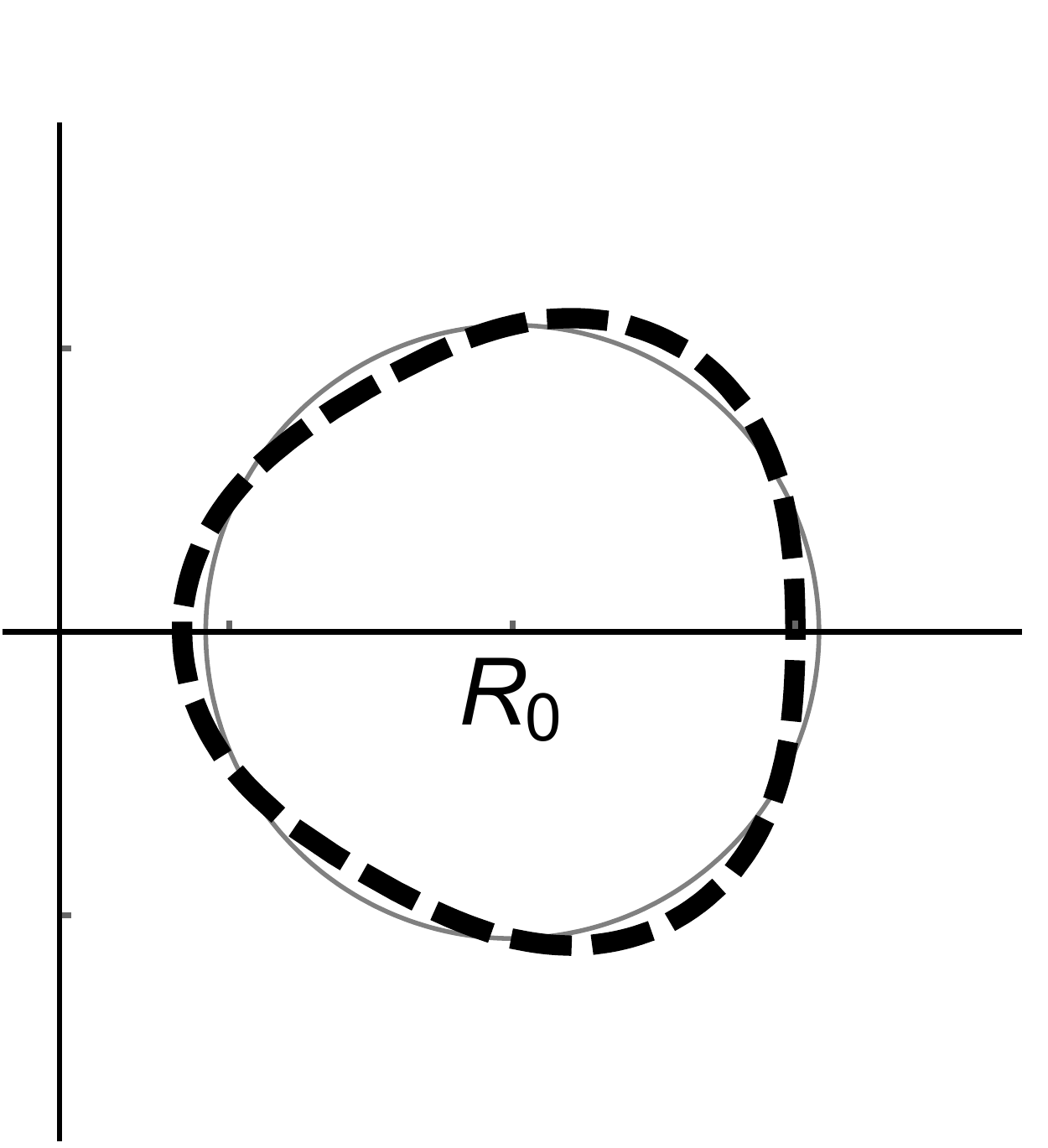}
 \includegraphics[width=0.18\textwidth]{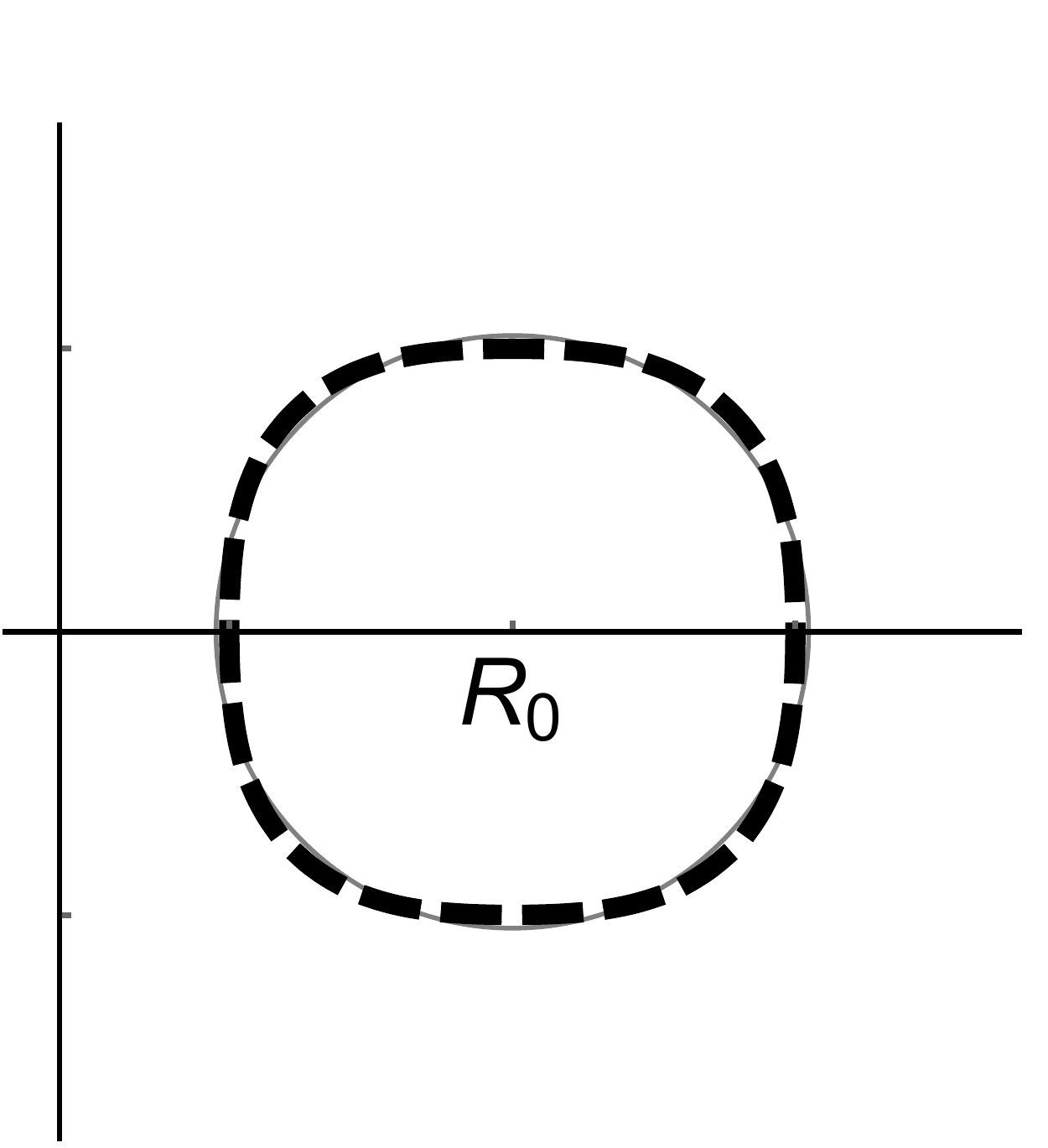}
 \includegraphics[width=0.18\textwidth]{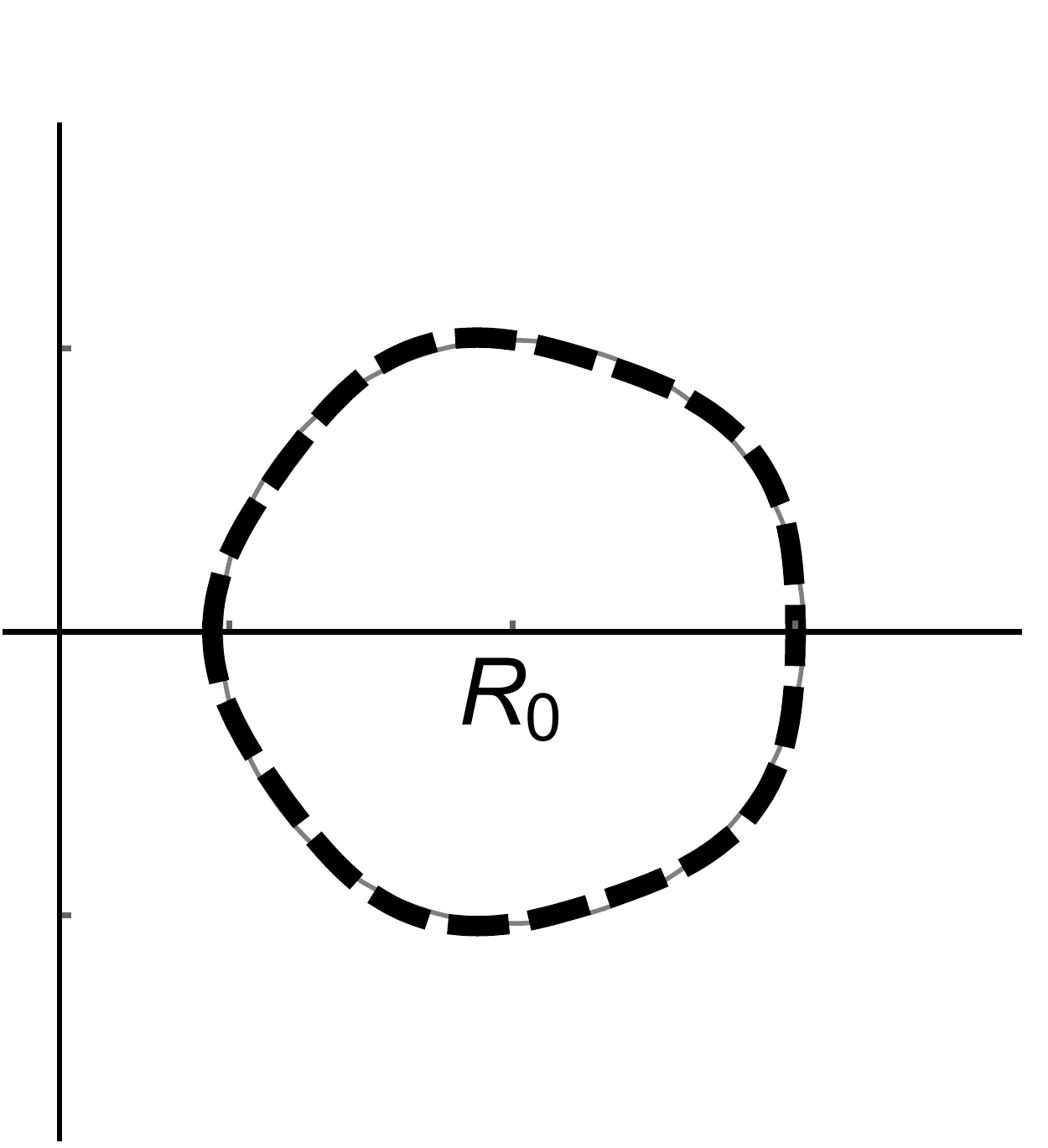}
 \includegraphics[width=0.18\textwidth]{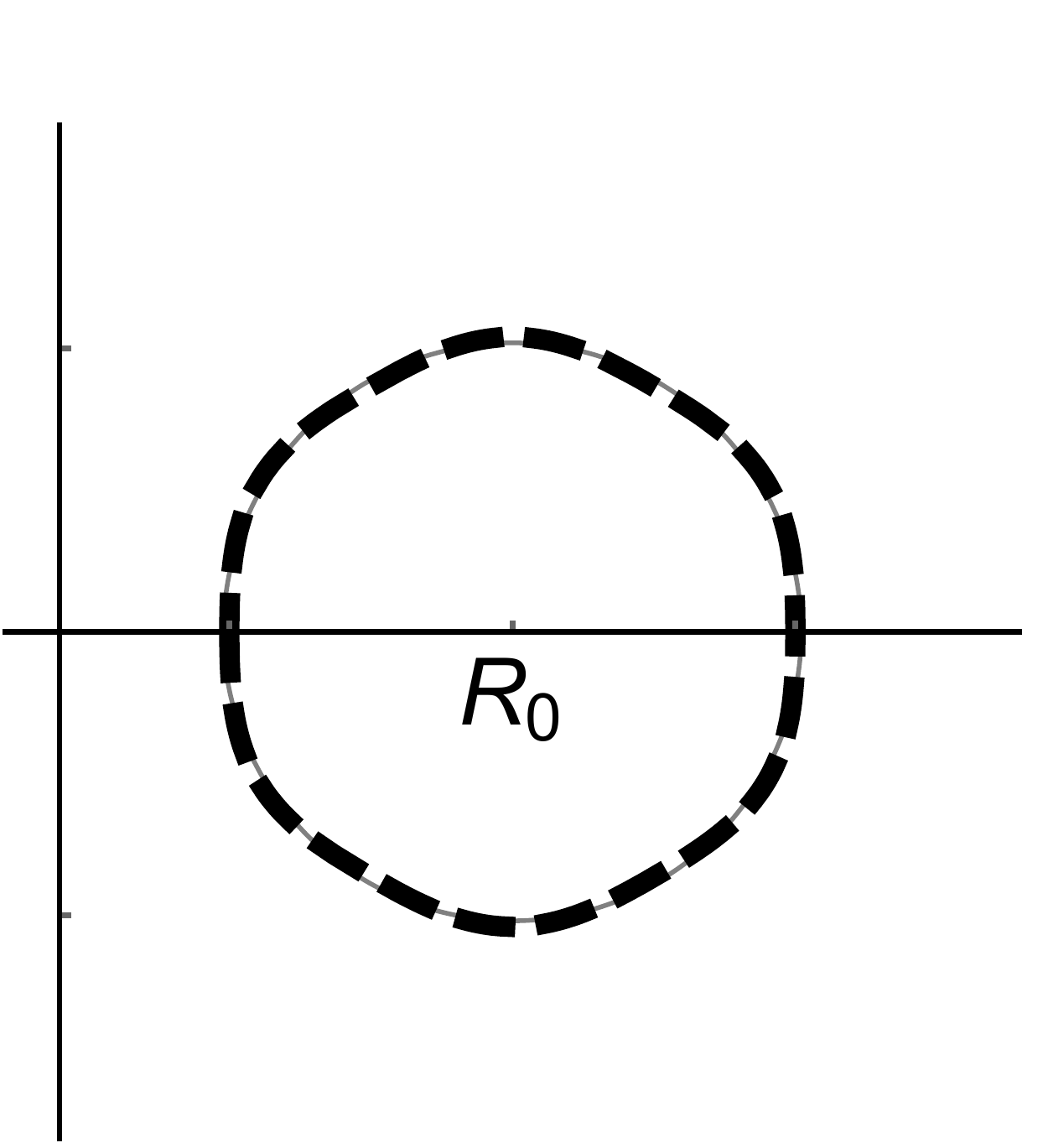}
 \includegraphics[width=0.0111\textwidth]{figs/xAxisLabelR.pdf}

 \caption{The $m_{c}=2$ through $m_{c}=6$ flux surface geometries in the tilted (solid) and up-down symmetric (dashed) configurations, with circular flux surfaces shown for comparison (gray).}
 \label{fig:simGeo}
\end{figure}

In section \ref{subsec:gyroSym}, we presented an analytic argument showing that (when expanding in $m_{c} \gg 1$) the solution to the gyrokinetic equation for a given geometry can be used to generate the solution to any geometry that is identical, except for a global tilt of the high order shaping effects. This relationship, given by equation \refEq{eq:symSolToAsymSol}, gives a prediction for the poloidal distributions of the fluxes. The full electromagnetic expressions are defined in \ref{app:fluxes}, but in the electrostatic limit they are given by
\begin{align}
  \gamma_{s}^{\phi} &\equiv - R \left\langle \left\langle \int d^{3} w \overline{h}_{s} \hat{e}_{\zeta} \cdot \delta \vec{E} \right\rangle_{\Delta \psi} \right\rangle_{\Delta t} \label{eq:polDistPartFlux} \\
  \pi_{s}^{\phi} &\equiv - R \left\langle \left\langle \int d^{3} w \overline{h}_{s} m_{s} R \left( \vec{w} \cdot \hat{e}_{\zeta} + R \Omega_{\zeta} \right) \hat{e}_{\zeta} \cdot \delta \vec{E} \right\rangle_{\Delta \psi} \right\rangle_{\Delta t} \label{eq:polDistMomFlux} \\
  q_{s}^{\phi} &\equiv - R \left\langle \left\langle \int d^{3} w \overline{h}_{s} \left( \frac{m_{s}}{2} w^{2} + Z_{s} e \Phi_{0} - \frac{m_{s}}{2} R^{2} \Omega_{\zeta}^{2} \right) \hat{e}_{\zeta} \cdot \delta \vec{E} \right\rangle_{\Delta \psi} \right\rangle_{\Delta t} \label{eq:polDistHeatFlux} \\
  p_{Q s}^{\phi} &\equiv \left\langle \left\langle \int d^{3} w Z_{s} e \overline{h}_{s} \frac{\partial \overline{\phi}}{\partial t} \right\rangle_{\Delta \psi} \right\rangle_{\Delta t} , \label{eq:polDistHeating}
\end{align}
which are just equations \refEq{eq:partFluxDef}, \refEq{eq:momFluxDef}, \refEq{eq:heatFluxDef}, and \refEq{eq:heatingDef} without the flux surface average (e.g. $Q_{s} = \left\langle q_{s} \right\rangle_{\psi}$). Specifically, using equation \refEq{eq:symSolToAsymSol} the analytic theory predicts that we should find
\begin{align}
  \gamma_{s}^{t} \left( \theta, z \right) &= \gamma_{s}^{u} \left( \theta, z + m_{c} \theta_{t} \right) \label{eq:symSolToAsymSolPart} \\
  \pi_{s}^{t} \left( \theta, z \right) &= \pi_{s}^{u} \left( \theta, z + m_{c} \theta_{t} \right) \label{eq:symSolToAsymSolMom} \\
  q_{s}^{t} \left( \theta, z \right) &= q_{s}^{u} \left( \theta, z + m_{c} \theta_{t} \right) , \\
  p_{Q s}^{t} \left( \theta, z \right) &= p_{Q s}^{u} \left( \theta, z + m_{c} \theta_{t} \right) , \label{eq:symSolToAsymSolHeat}
\end{align}
where the superscript $u$ indicates the geometry is up-down symmetric (i.e. untilted) and $t$ indicates the geometry is tilted. By simulating several geometries (see the up-down symmetric geometries shown in the bottom row of figure \ref{fig:simGeo}) and their corresponding tilted geometries (see the top row of figure \ref{fig:simGeo}) we can numerically verify equations \refEq{eq:symSolToAsymSolPart} through \refEq{eq:symSolToAsymSolHeat}. We will focus on the ion momentum flux because the symmetry has particularly profound consequences for it, but the analysis in this section can be applied to any of the fluxes.

We should note that GS2 automatically takes $\theta_{\alpha} \left( \psi \right) = 0$ in its definition of $\alpha$ (see equation \refEq{eq:alphaDef}), so we have to be careful about making numerical predictions from our analytic results. In general, converting between our definition of $\alpha$ and the GS2 definition, $\alpha_{\text{GS2}}$, involves accounting for a shift in $\alpha$ and $\vec{\nabla} \alpha$ of
\begin{align}
   \delta \left( \alpha \right) &= - I \left. \int_{\theta_{\alpha}}^{0} \right|_{\psi} d \theta' \left( R^{2} \vec{B} \cdot \vec{\nabla} \theta' \right)^{-1} \label{eq:GS2alphaShift} \\
   \delta \left( \vec{\nabla} \alpha \right) &= - I \left( \left. \int_{\theta_{\alpha}}^{0} \right|_{\psi} d \theta' F_{\alpha} \left( \theta' \right) - \left[ \frac{1}{R^{2} B_{p}} \left. \frac{\partial l_{p}}{\partial \theta} \right|_{\psi} \right]_{\theta = \theta_{\alpha}} \frac{d \theta_{\alpha}}{d \psi} \right) \vec{\nabla} \psi \label{eq:GS2gradAlphaShift}
\end{align}
respectively. However, given our specific choices in equations \refEq{eq:thetaAlphaDerivTilted} and \refEq{eq:thetaAlphaTilted} we see that
\begin{align}
   \delta \left( \alpha \right) &= 0 \\
   \delta \left( \vec{\nabla} \alpha \right) &= I \sum_{p = 1}^{\infty} \frac{\left( - 1 \right)^{p-1}}{m_{c}^{p}} \left( \Lambda^{p} \left[ \left. \frac{\partial^{p-1} F_{\alpha}}{\partial \theta^{p-1}} \right|_{z} \right] \left( 0, 0 \right) \right. \\
   &- \left. \Lambda^{p} \left[ \left. \frac{\partial^{p-1} F_{\alpha}}{\partial \theta^{p-1}} \right|_{z} \right] \left( 0, m_{c} \theta_{t} \right) \right) \vec{\nabla} \psi . \nonumber
\end{align}

The only effect of the shift in $\alpha$ is to introduce a phase factor of $\Exp{- i k_{\alpha} \delta \left( \alpha \right)}$ in the Fourier analyzed turbulent quantities $h_{s}$, $\phi$, $A_{||}$, and $B_{||}$ (e.g. equation \refEq{eq:distFnFourierAnalysis}). The shift in $\vec{\nabla} \alpha$ enters the gyrokinetic model only through
\begin{align}
   \vec{k}_{\perp} &= k_{\psi} \vec{\nabla} \psi + k_{\alpha} \vec{\nabla} \alpha = \left[ k_{\psi} + k_{\alpha} \frac{\partial \vec{r}}{\partial \psi} \cdot \delta \left( \vec{\nabla} \alpha \right) \right] \vec{\nabla} \psi + k_{\alpha} \vec{\nabla} \alpha_{\text{GS2}} . \label{eq:wavenumberShift}
\end{align}
Fortunately, neither of these changes has an effect on equations \refEq{eq:symSolToAsymSolPart} through \refEq{eq:symSolToAsymSolHeat}. The phase factor cancels because all transport is driven by the beating of two turbulent quantities (see \ref{app:fluxes}): one with the complex conjugate taken, the other without. As seen in equation \refEq{eq:wavenumberShift}, the tilt of $\vec{\nabla} \alpha$ can be taken into account by tilting $k_{\psi}$. Since the fluxes we are looking at involve the sum over all of wavenumber space, shifting flux from one wavenumber to another does not alter the total value.

\begin{figure}
 \begin{center}
  \includegraphics[width=0.5\textwidth]{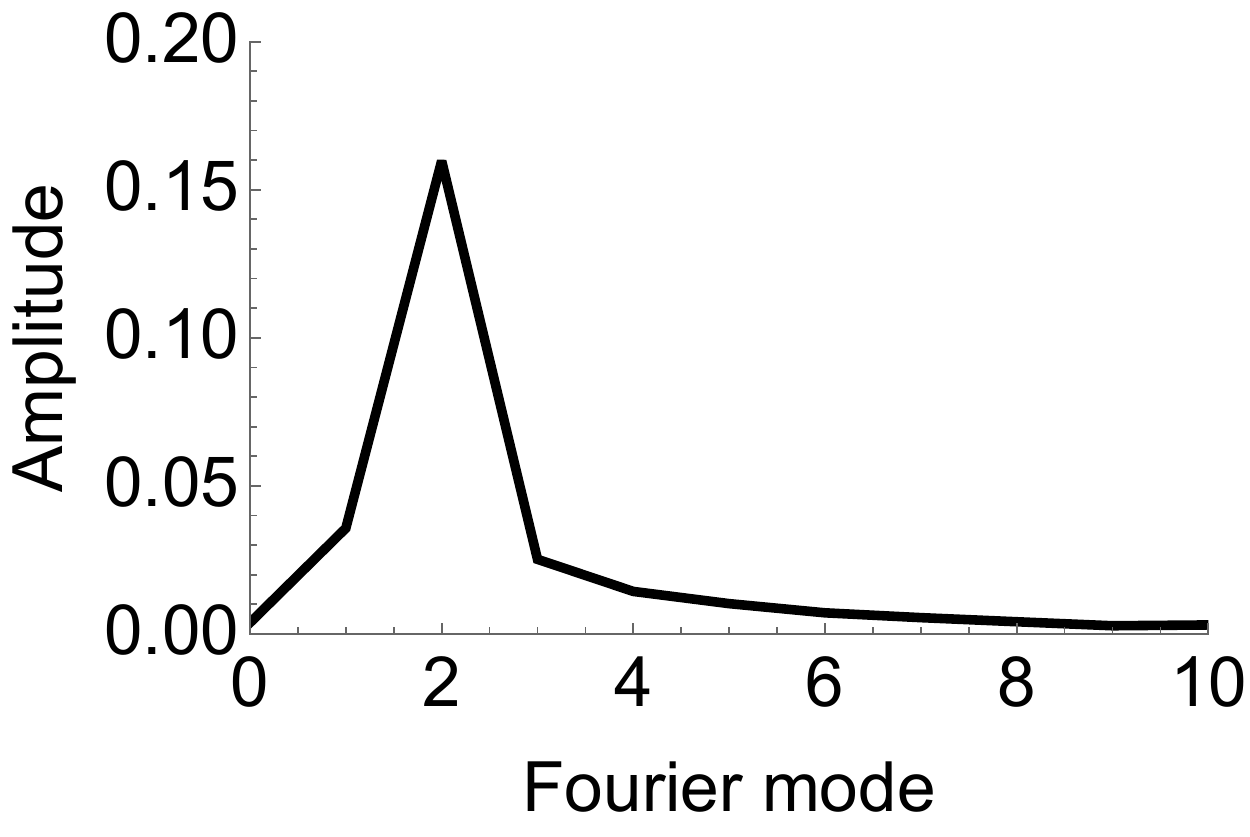}
 \end{center}
 \caption{The Fourier spectrum of the poloidal distribution of the ion momentum flux generated by circular flux surfaces.}
 \label{fig:circFourierSpectrum}
\end{figure}

Because GS2 is not constructed to separate the two spatial scales represented by $\theta$ and $z$, our simulations give $\pi_{s} \left( \theta \right) = \pi_{s} \left( \theta, z \left( \theta \right) \right)$ rather than $\pi_{s} \left( \theta, z \right)$. Therefore, we have to take the data produced by GS2, separate the dependences on the fast and slow poloidal coordinate, and then tilt only the fast spatial variation. We start by Fourier analyzing the poloidal distribution of momentum flux from GS2,
\begin{align}
   \pi_{s}^{u} \left( \theta \right) = \sum_{n = 1}^{\infty} P_{n} \Sin{n \theta} , \label{eq:GS2outputDist}
\end{align}
in the untilted case. Figure \ref{fig:circFourierSpectrum} shows a typical Fourier spectrum produced by GS2. Note that since the untilted case is up-down symmetric we know the momentum flux distribution must be odd \cite{ParraUpDownSym2011}, so we can neglect the $\Cos{n \theta}$ term. These even terms must be retained when considering the particle or heat fluxes. We want to transform equation \refEq{eq:GS2outputDist} into the form of a two dimensional Fourier series in the two separate spatial scales, e.g.
\begin{align}
   \pi_{s}^{u} \left( \theta, z \right) = \sum_{l = 0}^{\infty} \sum_{k = k_{\text{min}}}^{k_{\text{max}}} P_{k + l m_{c}} \Big( \Sin{l  z} \Cos{k \theta} + \Cos{l  z} \Sin{k \theta} \Big) . \label{eq:2dFourierSeries}
\end{align}
Using some trigonometric identities and equation \refEq{eq:zDef} it can be shown that if we choose to define $k$ as
\begin{align}
   k &\equiv n - l m_{c} , \label{eq:kDef}
\end{align}
then we can transform equation \refEq{eq:GS2outputDist} into equation \refEq{eq:2dFourierSeries} as along as $k_{max} - k_{min} = m_{c} - 1$.

\begin{figure}
 \begin{center}
  \includegraphics[width=0.5\textwidth]{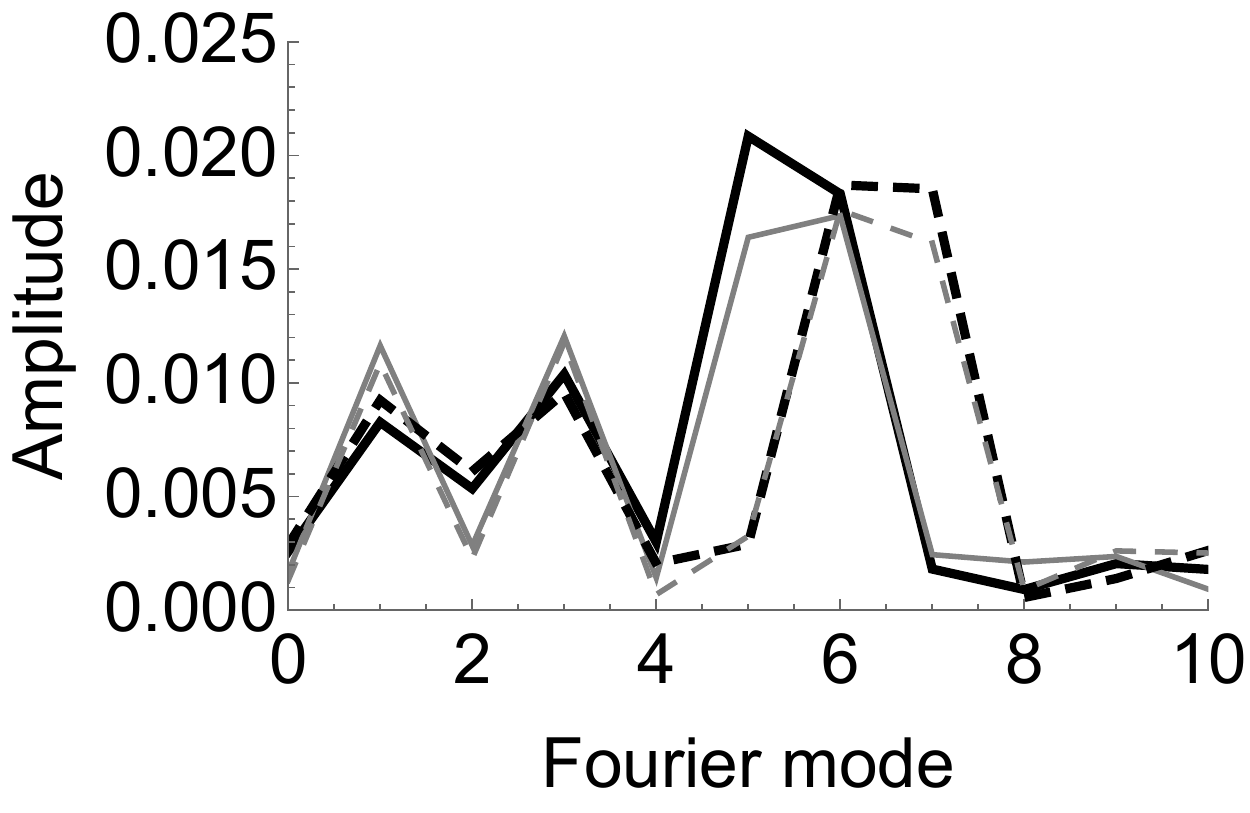}
 \end{center}
 \caption{The Fourier spectrum of the poloidal distribution of ion momentum flux after subtracting the flux generated by circular flux surfaces (shown in figure \ref{fig:circFourierSpectrum}) for up-down symmetric (gray) and tilted (black) configurations in the $m_{c} = 7$ (solid) and $m_{c} = 8$ (dashed) geometries.}
 \label{fig:m7m8FourierSpectrum}
\end{figure}

The definition of $l$ contains the physics of the scale separation and consequently will strongly affect how well we match GS2 results. The definition of $l$ controls which harmonics (enumerated by $n$) are mapped to $l = 0$ (and remain untilted), as opposed to $l = 1$ (which are tilted by $m_{c} \theta_{t}$), $l = 2$ (which are tilted by $2 m_{c} \theta_{t}$), etc. Intuitively we expect modes with $n \approx 1$ should remain untilted (i.e. map to $l = 0$), modes with $n \approx m_{c}$ should map to $l = 1$, and modes with $n \approx 2 m_{c}$ should map to $l = 2$. This general intuition motivates some sort of rounding to integers. The specific form of
\begin{align}
   l &\equiv \left\lfloor \frac{n + 2}{m_{c}} \right\rfloor \label{eq:lDef}
\end{align}
(where $\left\lfloor x \right\rfloor$ is the floor function that gives the integer value $n$ such that $n \leq x < n + 1$ for any real number $x$) was chosen in accordance with figure \ref{fig:m7m8FourierSpectrum}. We see that, as the shaping effect mode number $m_{c}$ is increased, the $m_{c} - 2$ and $m_{c} - 1$ Fourier terms of the momentum flux track with it, while all lower modes stay roughly constant. Unsurprisingly, this definition of $l$ was also found to produce the best agreement between theory and GS2 data. Our choice for $l$ means that $k_{\text{min}} = - 2$ and $k_{\text{max}} = m_{c} - 3$, leaving us with
\begin{align}
   \pi_{s}^{u} \left( \theta, z \right) = \sum_{l = 0}^{\infty} \sum_{k = - 2}^{m_{c} - 3} P_{k + l m_{c}} \Big( \Sin{l  z} \Cos{k \theta} + \Cos{l  z} \Sin{k \theta} \Big)
\end{align}
from equation \refEq{eq:2dFourierSeries}. Now we can use equation \refEq{eq:symSolToAsymSolMom} to construct
\begin{align}
   \pi_{s}^{t} \left( \theta, z \right) = \sum_{l = 0}^{\infty} \sum_{k = - 2}^{m_{c} - 3} P_{k + l m_{c}} \Big( \Sin{l  z + l m_{c} \theta_{t}} \Cos{k \theta} + \Cos{l  z + l m_{c} \theta_{t}} \Sin{k \theta} \Big) , \label{eq:tiltedMomPrediction}
\end{align}
a prediction for the distribution of momentum flux in the tilted geometry.

\begin{figure}
 \hspace{0.04\textwidth} (a) \hspace{0.4\textwidth} (b) \hspace{0.25\textwidth}
 \begin{center}
  \includegraphics[width=0.45\textwidth]{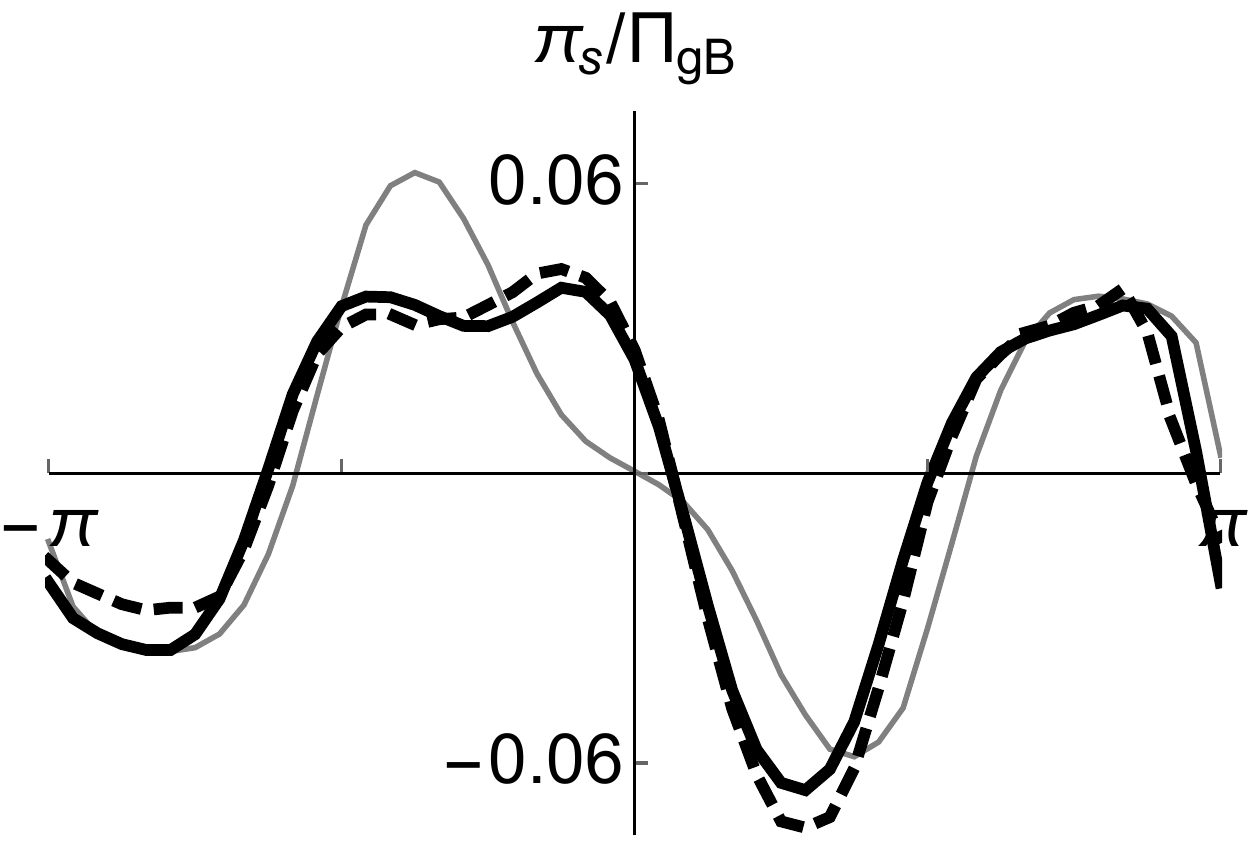}
  \includegraphics[width=0.45\textwidth]{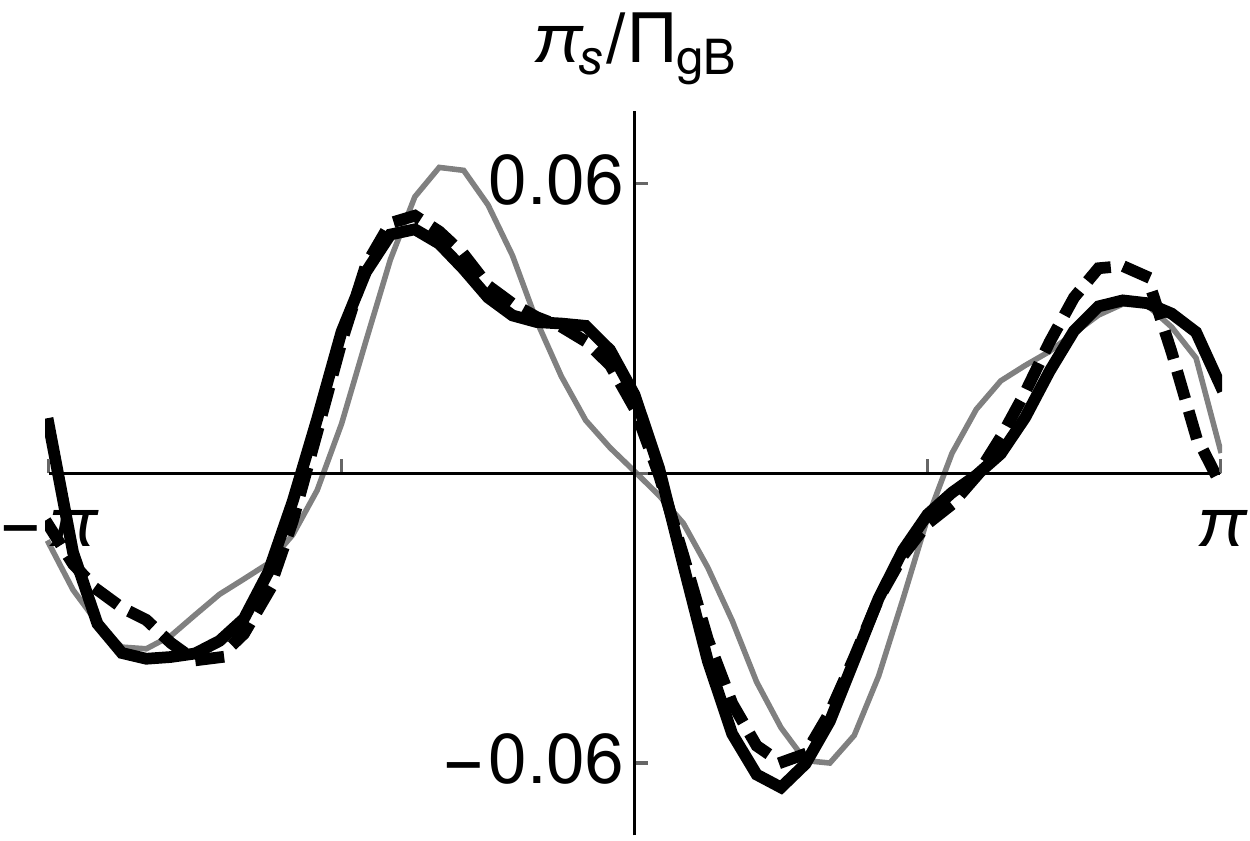}
  \raisebox{6.7\height}{$\theta$}
 \end{center}
 
 \hspace{0.04\textwidth} (c) \hspace{0.4\textwidth} (d) \hspace{0.25\textwidth}
 \begin{center}
  \includegraphics[width=0.45\textwidth]{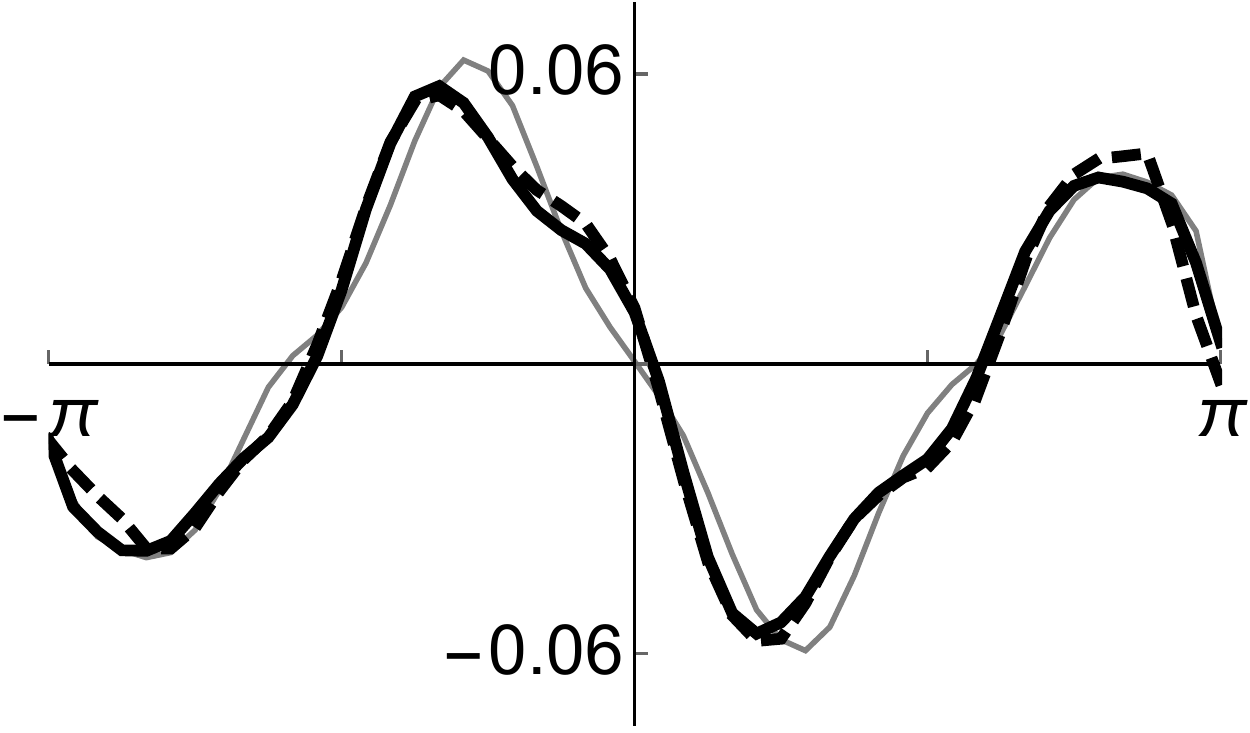}
  \includegraphics[width=0.45\textwidth]{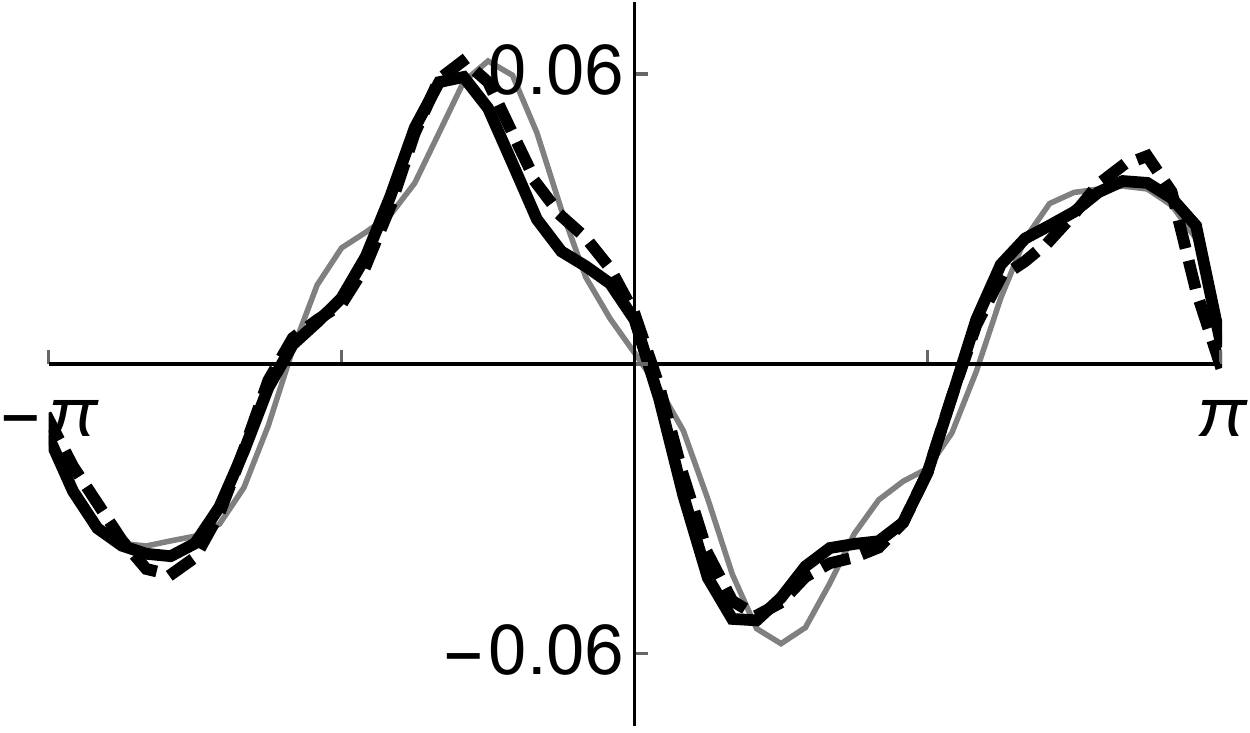}
  \raisebox{6.7\height}{$\theta$}
  \end{center}

 \caption{The full poloidal distribution of the ion momentum flux (see equation \refEq{eq:polDistMomFlux}) for the tilted geometry (black, thick), up-down symmetric geometry with the appropriate Fourier modes tilted (dashed, thick), and up-down symmetric geometry without any tilt (gray, thin), using (a) $m_{c} = 5$, (b) $m_{c} = 6$, (c) $m_{c} = 7$, and (d) $m_{c} = 8$ shaping modes (see figure \ref{fig:simGeo}). The momentum flux is normalized to the gyroBohm value.}
 \label{fig:fullMomProfiles}
\end{figure}

\begin{figure}
 \begin{center}
  \includegraphics[width=0.55\textwidth]{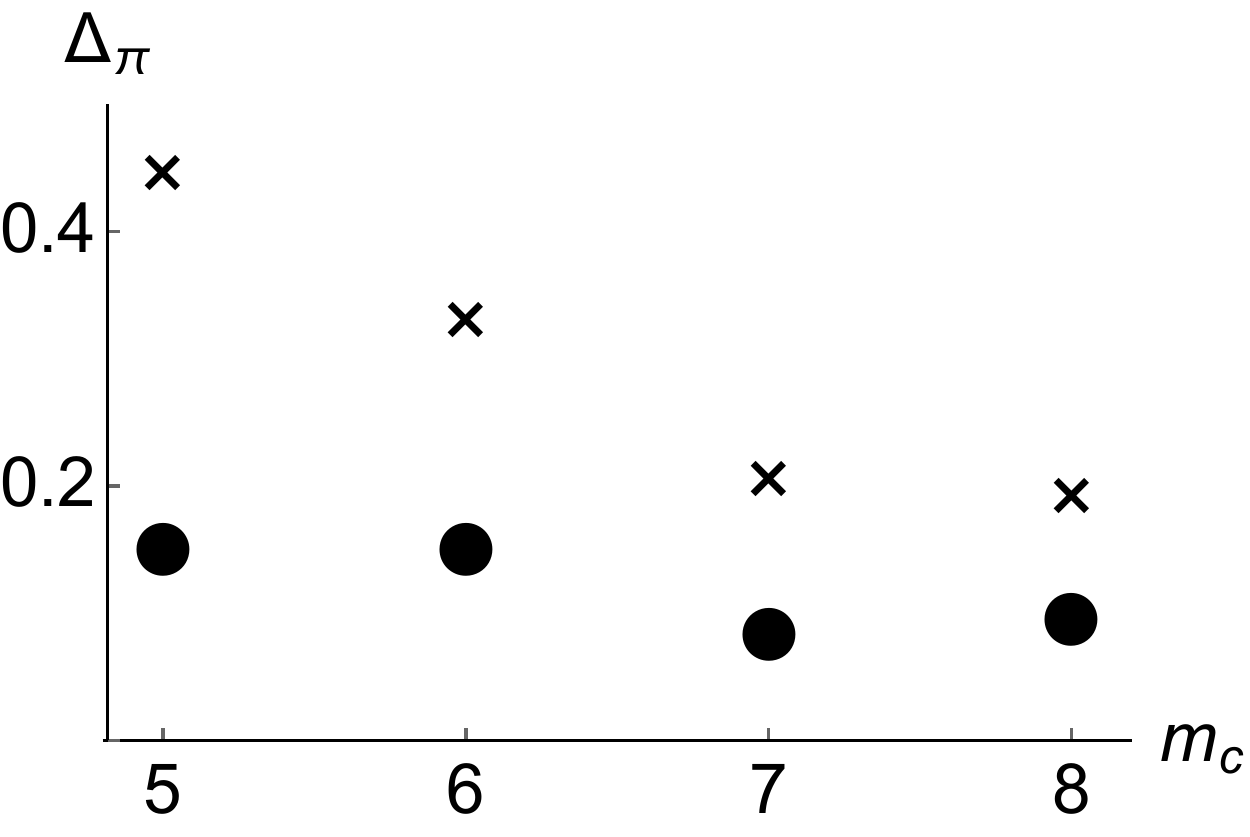}
 \end{center}
 \caption{The fractional error (see equation \refEq{eq:momDistFracError}) between the poloidal distribution of the momentum flux in the tilted geometry and the distribution predicted using the untilted geometry (circles), with the fractional error between the tilted and untilted (without any adjustment) shown as a control (crosses).}
 \label{fig:momDistFracError}
\end{figure}

Fundamentally, in this comparison we are testing the truth of equation \refEq{eq:symSolToAsymSol} and \refEq{eq:symSolToAsymSolMom}, which we used in deriving equation \refEq{eq:tiltedMomPrediction}. In figure \ref{fig:fullMomProfiles}, we use the numerical results from the untilted configuration and equation \refEq{eq:tiltedMomPrediction} to generate what we expect the momentum flux to be in the corresponding tilted configuration. Visually we see good agreement. In figure \ref{fig:momDistFracError} we quantify the agreement by calculating the fractional error according to
\begin{align}
   \Delta_{\pi} &\equiv \frac{\oint_{- \pi}^{\pi} d \theta \left| \pi_{s}^{\text{act}} \left( \theta \right) - \pi_{s}^{\text{calc}} \left( \theta \right) \right|}{\oint_{- \pi}^{\pi} d \theta \left| \pi_{s}^{\text{act}} \left( \theta \right) \right|} , \label{eq:momDistFracError}
\end{align}
where $\pi_{s}^{\text{act}} \left( \theta \right)$ is the momentum flux distribution from the tilted geometry (the thick black lines in figure \ref{fig:fullMomProfiles}) and $\pi_{s}^{\text{calc}} \left( \theta \right)$ is either the predicted value calculated from the untilted geometry (the dashed black lines in figure \ref{fig:fullMomProfiles}) or the raw untilted geometry (the solid gray lines in figure \ref{fig:fullMomProfiles}) to serve as a control. As is also apparent from figure \ref{fig:fullMomProfiles}, when we look at geometries with larger values of $m_{c}$ we find better agreement between the titled geometry and the up-down symmetric geometry (when appropriately translated). The agreement breaks down significantly below $m_{c} = 5$ and we have enough information to understand why. Extrapolating from figure \ref{fig:m7m8FourierSpectrum}, we would expect $m_{c} = 4$ shaping to be problematic because it generates an ion momentum flux distribution with a strong Fourier mode at $2$. This would be indistinguishable from the dominant Fourier mode at $2$, which is inherent to tokamaks (see figure \ref{fig:circFourierSpectrum}). Since one cannot separate these two contributions (the one from the $m_{c} = 4$ shaping and the one inherent to tokamaks), it is not possible to translate the contribution from the $m_{c} = 4$ shaping as is appropriate.

These numerical results verify equation \refEq{eq:symSolToAsymSol} and the derivation of section \ref{subsec:gyroSym}. Additionally, though not shown here, the poloidal distributions of particle, momentum, and heat flux (for both ions and electrons) all agree with theory in a similar manner to what is seen in figure \ref{fig:fullMomProfiles}.

%===================================================%
\subsection{Change in total fluxes with tilt}
\label{subsec:totalFluxComp}
%===================================================%

\begin{figure}
 \begin{center}
  \includegraphics[width=0.7\textwidth]{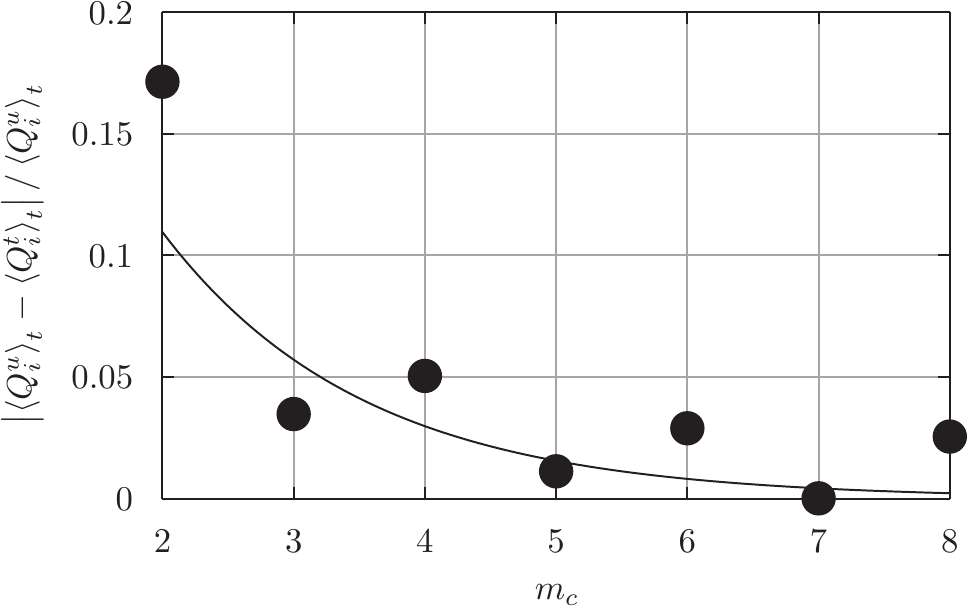}
 \end{center}
 \caption{The fractional difference in the electrostatic ion heat flux between up-down symmetric and tilted geometries (see figure \ref{fig:simGeo}) as a function of $m_{c}$ (points), with an exponential fit of the form $K \Exp{- \beta m_{c}}$ (line).}
 \label{fig:heatFluxChange}
\end{figure}

From section \ref{subsec:gyroSym} we expect the change in the turbulent fluxes (see equations \refEq{eq:partFluxAvg} through \refEq{eq:heatFluxAvg}) due to tilting the fast shaping effects to be exponentially small in $m_{c} \gg 1$. In figure \ref{fig:heatFluxChange}, we show the fractional difference between the ion heat flux from an up-down symmetric configuration and the corresponding tilted configuration, for the geometry of figure \ref{fig:simGeo}. We see that the difference is consistent with an exponential as expected. The difference is pronounced for the $m_{c} = 2$ case and rapidly diminishes at higher $m_{c}$.

%===================================================%
%===================================================%
\section{Conclusions}
\label{sec:conclusions}
%===================================================%
%===================================================%

The poloidal tilting symmetry shown in this work demonstrates that the tilt angle of high order flux surface shaping has little effect on transport of particles, momentum, or energy. This suggests that tilting elongation will have a larger effect on transport than tilting higher order modes (e.g. triangularity or squareness). Additionally the tilting symmetry establishes a close connection between up-down symmetric devices and devices that have flux surfaces with mirror symmetry. This is because all flux surfaces that have mirror symmetry can be produced by tilting a corresponding up-down symmetric flux surface by a single tilt angle. This correspondence distinguishes mirror symmetric devices from non-mirror symmetric devices, which has particular significance for momentum transport because up-down symmetric devices do not transport momentum to lowest order in $\rho_{\ast} \equiv \rho_{i} / a \ll 1$. Therefore, breaking the mirror symmetry of flux surfaces may increase the momentum transport generated by up-down asymmetry.

\ack

This work was funded in part by the RCUK Energy Programme (grant number EP/I501045). Computing time for this work was provided by the Helios supercomputer at IFERC-CSC under the projects SPIN, MULTEIM, and GKMSC. The authors also acknowledge the use of ARCHER through the Plasma HEC Consortium EPSRC grant number EP/L000237/1 under the projects e281-gs2 and e281-rotation.

\appendix

%===================================================%
%===================================================%
\section{Electromagnetic turbulent fluxes and heating}
\label{app:fluxes}
%===================================================%
%===================================================%

From references \cite{ParraUpDownSym2011, SugamaHighFlowGyro1998, AbelGyrokineticsDeriv2012} among others we see that the turbulent electromagnetic fluxes of particles, momentum, and energy as well as the turbulent energy exchange between species are the only turbulent quantities needed to evolve the transport equations for particles, momentum, and energy. Furthermore, it is convenient to calculate these fluxes in a frame rotating with the bulk plasma, using the velocity variable $\vec{w} \equiv \vec{v} - R \Omega_{\zeta} \hat{e}_{\zeta}$. To do so we will follow the procedure outlined in Section II.D and Appendix E of reference \cite{ParraUpDownSym2011}.

The complete electromagnetic turbulent flux of particles in a tokamak can be defined as
\begin{align}
   \Gamma_{s} &\equiv \left\langle \gamma_{s} \right\rangle_{\psi} \equiv - \left\langle R \left\langle \left\langle \int d^{3} w \overline{h}_{s} \hat{e}_{\zeta} \cdot \left( \delta \vec{E} + \vec{w} \times \delta \vec{B} \right) \right\rangle_{\Delta \psi} \right\rangle_{\Delta t} \right\rangle_{\psi} ,
\end{align}
where $\gamma_{s}$ is the poloidally-dependent particle flux, $\delta \vec{E} = - \vec{\nabla}_{\perp} \overline{\phi}$ is the turbulent electric field, and $\delta \vec{B} = \overline{B}_{||} \hat{b} + \vec{\nabla} \overline{A}_{||} \times \hat{b}$ is the turbulent magnetic field. After considerable manipulation we find the flux of particles to be
\begin{align}
   \Gamma_{s} =& \frac{4 \pi^{2} i}{m_{s} V'} \left\langle \sum_{k_{\psi}, k_{\alpha}} k_{\alpha} \oint d \theta J B \int dw_{||} d \mu ~ h_{s} \left( - k_{\psi}, - k_{\alpha} \right) \right. \nonumber \\
   & \times \Bigg[ \phi \left( k_{\psi}, k_{\alpha} \right) J_{0} \left( k_{\perp} \rho_{s} \right) \\
   &- A_{||} \left( k_{\psi}, k_{\alpha} \right) w_{||} J_{0} \left( k_{\perp} \rho_{s} \right) \nonumber \\
   &+ \left. B_{||} \left( k_{\psi}, k_{\alpha} \right) \frac{1}{\Omega_{s}} \frac{\mu B}{m_{s}} \frac{2 J_{1} \left( k_{\perp} \rho_{s} \right)}{k_{\perp} \rho_{s}} \Bigg] \right\rangle_{\Delta t} . \nonumber
\end{align}

The complete electromagnetic turbulent flux of toroidal angular momentum in a tokamak can be defined as
\begin{align}
   \Pi &\equiv \sum_{s} \Pi_{s} + \Pi_{B} ,
\end{align}
where
\begin{align}
   \Pi_{s} &\equiv \left\langle \pi_{s} \right\rangle_{\psi} \equiv - \left\langle R \left\langle \left\langle \int d^{3} w \overline{h}_{s} m_{s} R \left( \vec{w} \cdot \hat{e}_{\zeta} + R \Omega_{\zeta} \right) \hat{e}_{\zeta} \cdot \left( \delta \vec{E} + \vec{w} \times \delta \vec{B} \right) \right\rangle_{\Delta \psi} \right\rangle_{\Delta t} \right\rangle_{\psi}
\end{align}
is the contribution from particles,
\begin{align}
   \Pi_{B} &\equiv - \left\langle R \left\langle \left\langle \hat{e}_{\zeta} \cdot \tensor{\sigma} \cdot \vec{\nabla} \psi \right\rangle_{\Delta \psi} \right\rangle_{\Delta t} \right\rangle_{\psi}
\end{align}
is the momentum transported by the electromagnetic fields, $\pi_{s}$ is the poloidally-dependent angular momentum flux,
\begin{align}
   \tensor{\sigma} \equiv \frac{1}{\mu_{0}} \vec{B} \vec{B} - \frac{1}{2 \mu_{0}} B^{2} \tensor{I}
\end{align}
is the Maxwell stress tensor, and $\tensor{I}$ is the identity matrix. After considerable manipulation we find the angular momentum transported by particles to be
\begin{align}
   \Pi_{s} =& \frac{4 \pi^{2} i}{V'} \left\langle \sum_{k_{\psi}, k_{\alpha}} k_{\alpha} \oint d \theta J B \int dw_{||} d \mu ~ h_{s} \left( - k_{\psi}, - k_{\alpha} \right) \right. \nonumber \\
   & \times \left\{ \phi \left( k_{\psi}, k_{\alpha} \right) \left[ \left( \frac{I}{B} w_{||} + R^{2} \Omega_{\zeta} \right) J_{0} \left( k_{\perp} \rho_{s} \right) + \frac{i}{\Omega_{s}} \frac{k^{\psi}}{B} \frac{\mu B}{m_{s}} \frac{2 J_{1} \left( k_{\perp} \rho_{s} \right)}{k_{\perp} \rho_{s}} \right] \right. \label{eq:speciesMomFlux} \\
   &- A_{||} \left( k_{\psi}, k_{\alpha} \right) \left[ \left( \frac{I}{B} w_{||} + R^{2} \Omega_{\zeta} \right) w_{||} J_{0} \left( k_{\perp} \rho_{s} \right) + \left( i \frac{w_{||}}{\Omega_{s}} \frac{k^{\psi}}{B} + \frac{I}{B} \right) \frac{\mu B}{m_{s}} \frac{2 J_{1} \left( k_{\perp} \rho_{s} \right)}{k_{\perp} \rho_{s}} \right] \nonumber \\
   &+ \left. \left. B_{||} \left( k_{\psi}, k_{\alpha} \right) \frac{1}{\Omega_{s}} \left[ \left( \frac{I}{B} w_{||} + R^{2} \Omega_{\zeta} \right) \frac{\mu B}{m_{s}} \frac{2 J_{1} \left( k_{\perp} \rho_{s} \right)}{k_{\perp} \rho_{s}} + \frac{i}{2 \Omega_{s}} \frac{k^{\psi}}{B} \frac{\mu^{2} B^{2}}{m_{s}^{2}} G \left( k_{\perp} \rho_{s} \right) \right] \right\} \right\rangle_{\Delta t} . \nonumber
\end{align}
and the transport by the fluctuating fields to be
\begin{align}
   \Pi_{B} &= \frac{2 \pi i}{\mu_{0} V'} \left\langle \sum_{k_{\psi}, k_{\alpha}} k_{\alpha} \oint d \theta J A_{||} \left( k_{\psi}, k_{\alpha} \right) \Big[ - i k^{\psi} A_{||} \left( - k_{\psi}, - k_{\alpha} \right) + I B_{||} \left( - k_{\psi}, - k_{\alpha} \right) \Big] \right\rangle_{\Delta t} , \label{eq:elecMagMomFlux}
\end{align}
where $k^{\psi} \equiv \vec{k}_{\perp} \cdot \vec{\nabla} \psi = k_{\psi} \left| \vec{\nabla} \psi \right|^{2} + k_{\alpha} \vec{\nabla} \psi \cdot \vec{\nabla} \alpha$ and $G \left( x \right) \equiv 8 \left( 2 J_{1} \left( x \right) - x J_{0} \left( x \right) \right) / x^{3}$. Note that, when summing over all species, equation \refEq{eq:perpCur} can be used to show that the $B_{||}$ term in equation \refEq{eq:elecMagMomFlux} cancels the fourth $A_{||}$ term in equation \refEq{eq:speciesMomFlux}.

The complete electromagnetic turbulent flux of energy carried by particles can be defined as
\begin{align}
   Q_{s} &\equiv \left\langle q_{s} \right\rangle_{\psi} \equiv - \left\langle R \left\langle \left\langle \int d^{3} w \overline{h}_{s} \left( \frac{m_{s}}{2} w^{2} + Z_{s} e \Phi_{0} - \frac{m_{s}}{2} R^{2} \Omega_{\zeta}^{2} \right) \right. \right. \right. \\
   & \hat{e}_{\zeta} \cdot \left. \left. \left. \left( \delta \vec{E} + \vec{w} \times \delta \vec{B} \right) \right\rangle_{\Delta \psi} \right\rangle_{\Delta t} \right\rangle_{\psi} , \nonumber
\end{align}
where $q_{s}$ is the poloidally-dependent energy flux. After considerable manipulation we find the energy transported by particles to be
\begin{align}
   Q_{s} =& \frac{4 \pi^{2} i}{V'} \left\langle \sum_{k_{\psi}, k_{\alpha}} k_{\alpha} \oint d \theta J B \int dw_{||} d \mu ~ h_{s} \left( - k_{\psi}, - k_{\alpha} \right) \left( \frac{w^{2}}{2} + \frac{Z_{s} e \Phi_{0}}{m_{s}} - \frac{m_{s}}{2} R^{2} \Omega_{\zeta}^{2} \right) \right. \nonumber \\
   & \times \left[ \phi \left( k_{\psi}, k_{\alpha} \right) \left( J_{0} \left( k_{\perp} \rho_{s} \right) \right) \right. \label{eq:speciesHeatFlux} \\
   &- A_{||} \left( k_{\psi}, k_{\alpha} \right) \left( w_{||} J_{0} \left( k_{\perp} \rho_{s} \right) \right) \nonumber \\
   &+ \left. \left. B_{||} \left( k_{\psi}, k_{\alpha} \right) \frac{1}{\Omega_{s}} \left( \frac{\mu B}{m_{s}} \frac{2 J_{1} \left( k_{\perp} \rho_{s} \right)}{k_{\perp} \rho_{s}} \right) \right] \right\rangle_{\Delta t} . \nonumber
\end{align}

The complete electromagnetic turbulent energy exchange between species can be written as
\begin{align}
   P_{Q s} &\equiv \left\langle p_{Q s} \right\rangle_{\psi} \equiv \left\langle \left\langle \left\langle \int d^{3} w Z_{s} e \overline{h}_{s} \frac{\partial \overline{\chi}}{\partial t} \right\rangle_{\Delta \psi} \right\rangle_{\Delta t} \right\rangle_{\psi} ,
\end{align}
where $\overline{\chi} \equiv \overline{\phi} - \vec{w} \cdot \vec{\overline{A}}$ is the generalized potential. After considerable manipulation we find the energy exchange to be
\begin{align}
   P_{Q s} =& \frac{4 \pi^{2}}{V'} \left\langle \sum_{k_{\psi}, k_{\alpha}} \oint d \theta J \Omega_{s} \int dw_{||} d \mu ~ h_{s} \left( - k_{\psi}, - k_{\alpha} \right) \right. \nonumber \\
   & \times \Bigg[ \frac{\partial}{\partial t} \left( \phi \left( k_{\psi}, k_{\alpha} \right) \right) J_{0} \left( k_{\perp} \rho_{s} \right) \\
   &- \frac{\partial}{\partial t} \left( A_{||} \left( k_{\psi}, k_{\alpha} \right) \right) w_{||} J_{0} \left( k_{\perp} \rho_{s} \right) \nonumber \\
   &+ \left. \frac{\partial}{\partial t} \left( B_{||} \left( k_{\psi}, k_{\alpha} \right) \right) \frac{1}{\Omega_{s}} \frac{\mu B}{m_{s}} \frac{2 J_{1} \left( k_{\perp} \rho_{s} \right)}{k_{\perp} \rho_{s}} \Bigg] \right\rangle_{\Delta t} . \nonumber
\end{align}

%===================================================%
%===================================================%
\section{Alternative calculation for $\vec{\nabla} \alpha$ integral}
\label{app:alphaIntegral}
%===================================================%
%===================================================%

Here we will show that, when you tilt the fast shaping (i.e. $z \left( \theta \right) = z_{t} \left( \theta \right) = m_{c} \left( \theta + \theta_{t} \right)$) of a given geometry, it has no effect on the integral appearing in $\vec{\nabla} \alpha$, except by modifying the form of $z \left( \theta \right)$. This was demonstrated in section \ref{subsec:gyroSym}, but here we present an alternative method. To address the integral appearing in equations \refEq{eq:QgeoSimple} and \refEq{eq:GalphaThetaDef} we will first choose the free parameter $\theta_{\alpha} \left( \psi \right)$ in the untilted case such that it and its radial derivative vanish on the flux surface of interest (i.e. $\theta_{\alpha} = 0$ and $d \theta_{\alpha} / d \psi = 0$). Then we can define
\begin{align}
   G_{\alpha u}^{\theta} \left( \theta \right) \equiv G_{\alpha u}^{\theta} \left( \theta, z_{u} \left( \theta \right) \right) = \left. \int_{0}^{\theta} \right|_{\psi} d \theta' F_{\alpha} \left( \theta', z_{u} \left( \theta' \right) \right) \label{eq:GalphaThetaDefUntilted}
\end{align}
by equation \refEq{eq:GalphaThetaDef}, which only depends on $\theta$. However, we can reintroduce the fast spatial scale by first Fourier analyzing in $\theta$ to get
\begin{align}
   G_{\alpha u}^{\theta} \left( \theta \right) = P_{\text{shear}} \theta + \sum_{n = 0}^{\infty} \Big( P^{S}_{n} \Sin{n \theta} + P^{C}_{n} \Cos{n \theta} \Big) .
\end{align}
We know that $G_{\alpha u}^{\theta}$ must have this form because $F_{\alpha}$ is periodic in both $\theta$ and $z$ (see equation \refEq{eq:FalphaDef}), but can have a poloidal average value due to the magnetic shear. Then we can rearrange, introducing $n = k + l m_{c}$ and using some trigonometric identities, to get
\begin{align}
   G_{\alpha u}^{\theta} \left( \theta \right) =& P_{\text{shear}} \theta + \sum_{l = 0}^{\infty} \sum_{k = k_{\text{min}}}^{k_{\text{max}}} \left[ P^{S}_{k + l m_{c}} \Big( \Sin{l m_{c} \theta} \Cos{k \theta} + \Cos{l m_{c} \theta} \Sin{k \theta} \Big) \right. \nonumber \\
   &+ \left. P^{C}_{k + l m_{c}} \Big( \Cos{l m_{c} \theta} \Cos{k \theta} - \Sin{l m_{c} \theta} \Sin{k \theta} \Big) \right] .
\end{align}
Note that this result is a generalization of equation \refEq{eq:2dFourierSeries} to include even terms and the secular linear term. For this form to be physically meaningful we rely on having a clear separation of scales such that $k_{\text{min}} \ll m_{c}$ and $k_{\text{max}} \ll m_{c}$. Now we can define
\begin{align}
   G_{\alpha} \left( \theta, z \right) \equiv& P_{\text{shear}} \theta + \sum_{l = 0}^{\infty} \sum_{k = k_{\text{min}}}^{k_{\text{max}}} \left[ P^{S}_{k + l m_{c}} \Big( \Sin{l z} \Cos{k \theta} + \Cos{l z} \Sin{k \theta} \Big) \right. \nonumber \\
   &+ \left. P^{C}_{k + l m_{c}} \Big( \Cos{l z} \Cos{k \theta} - \Sin{l z} \Sin{k \theta} \Big) \right] \label{eq:GalphaDef}
\end{align}
and check that, by substituting in $z = z_{u} \left( \theta \right) = m_{c} \theta$, our definition satisfies
\begin{align}
   G_{\alpha} \left( \theta, z_{u} \left( \theta \right) \right) = G_{\alpha u}^{\theta} \left( \theta \right) . \label{eq:sepScaleEquiv}
\end{align}
Using this result, equation \refEq{eq:GalphaThetaDefUntilted}, and our choice that $\theta_{\alpha} = 0$ and $d \theta_{\alpha} / d \psi = 0$ for the untilted case, we see that
\begin{align}
   \left. \int_{\theta_{\alpha}}^{\theta} \right|_{\psi} d \theta' F_{\alpha} \left( \theta', z_{u} \left( \theta' \right) \right) - \left[ \frac{1}{R^{2} B_{p}^{2}} \left. \frac{\partial l_{p}}{\partial \theta} \right|_{\psi} \right]_{\theta = \theta_{\alpha}} \frac{d \theta_{\alpha}}{d \psi} = G_{\alpha} \left( \theta, z_{u} \left( \theta \right) \right) . \label{eq:untiltedAlphaResult}
\end{align}

Next, by the fundamental theorem of calculus, equations \refEq{eq:GalphaThetaDefUntilted} and \refEq{eq:sepScaleEquiv} imply
\begin{align}
   \frac{d}{d \theta} \Big( G_{\alpha} \left( \theta, z_{u} \left( \theta \right) \right) \Big) = F_{\alpha} \left( \theta, z_{u} \left( \theta \right) \right) . \label{eq:GalphaDeriv}
\end{align}
However we also see that
\begin{align}
   \frac{d}{d \theta} \Big( G_{\alpha} \left( \theta, z_{u} \left( \theta \right) \right) \Big) &= \left. \frac{\partial G_{\alpha}}{\partial \theta} \right|_{z_{u}} + m_{c} \left. \frac{\partial G_{\alpha}}{\partial z_{u}} \right|_{\theta} \label{eq:untiltedGalphaChainRule} \\
   \frac{d}{d \theta} \Big( G_{\alpha} \left( \theta, z_{t} \left( \theta \right) \right) \Big) &= \left. \frac{\partial G_{\alpha}}{\partial \theta} \right|_{z_{t}} + m_{c} \left. \frac{\partial G_{\alpha}}{\partial z_{t}} \right|_{\theta} . \label{eq:tiltedGalphaChainRule}
\end{align}
Because the derivatives of $z_{u} \left( \theta \right)$ and $z_{t} \left( \theta \right)$ are identical, equations \refEq{eq:untiltedGalphaChainRule} and \refEq{eq:tiltedGalphaChainRule} have the same form. This allows us to use equation \refEq{eq:GalphaDeriv} to establish that
\begin{align}
   \frac{d}{d \theta} \Big( G_{\alpha} \left( \theta, z_{t} \left( \theta \right) \right) \Big) = F_{\alpha} \left( \theta, z_{t} \left( \theta \right) \right) .
\end{align}
Again using the fundamental theorem of calculus, we find
\begin{align}
   \left. \int_{\theta_{\alpha}}^{\theta} \right|_{\psi} d \theta' F_{\alpha} \left( \theta', z_{t} \left( \theta' \right) \right) = G_{\alpha} \left( \theta, z_{t} \left( \theta \right) \right) - G_{\alpha} \left( \theta_{\alpha}, z_{t} \left( \theta_{\alpha} \right) \right) .
\end{align}
Therefore, we will carefully select the free parameter $\theta_{\alpha} \left( \psi \right)$ in the tilted case so that
\begin{align}
   \left. \int_{\theta_{\alpha}}^{\theta} \right|_{\psi} d \theta' F_{\alpha} \left( \theta', z_{t} \left( \theta' \right) \right) - \left[ \frac{1}{R^{2} B_{p}^{2}} \left. \frac{\partial l_{p}}{\partial \theta} \right|_{\psi} \right]_{\theta = \theta_{\alpha}} \frac{d \theta_{\alpha}}{d \psi} = G_{\alpha} \left( \theta, z_{t} \left( \theta \right) \right) . \label{eq:tiltedAlphaResult}
\end{align}
This is done by choosing
\begin{align}
   \theta_{\alpha} &= 0 \label{eq:thetaAlphaChoice} \\
   \frac{d \theta_{\alpha}}{d \psi} &= - \left[ \frac{1}{R^{2} B_{p}^{2}} \left. \frac{\partial l_{p}}{\partial \theta} \right|_{\psi} \right]_{\theta = 0}^{-1} G_{\alpha} \left( 0, z_{t} \left( 0 \right) \right) . \label{eq:thetaAlphaDerivChoice}
\end{align}
Note that this is identical to the results of the main text (see equations \refEq{eq:thetaAlphaDerivTilted} and \refEq{eq:thetaAlphaTilted}) because
\begin{align}
   G_{\alpha} \left( 0, z_{t} \left( 0 \right) \right) &= - \sum_{p = 1}^{\infty} \frac{\left( - 1 \right)^{p-1}}{m_{c}^{p}} \left( \Lambda^{p} \left[ \left. \frac{\partial^{p-1} F_{\alpha}}{\partial \theta^{p-1}} \right|_{z} \right] \left( 0, 0 \right) - \Lambda^{p} \left[ \left. \frac{\partial^{p-1} F_{\alpha}}{\partial \theta^{p-1}} \right|_{z} \right] \left( 0, m_{c} \theta_{t}\right) \right) .
\end{align}

We see by comparing equation \refEq{eq:untiltedAlphaResult} for the untilted case with equation \refEq{eq:tiltedAlphaResult} for the tilted case that the effect of the tilt can be entirely contained in the form of $z \left( \theta \right)$. This means that the geometric coefficients for both the untilted and tilted cases can be written in the form $Q_{\text{geo}} \left( \theta, z \right)$, where $z = z_{u}$ for the untilted case and $z = z_{t}$ for the tilted case.

%===================================================%
%===================================================%
\section*{References}
\bibliographystyle{unsrt}
%\bibliography{/Users/Justin/Documents/Research/Bibliography/references.bib}
\bibliography{references.bib}
%===================================================%
%===================================================%

\end{document}